\documentclass[aps,prd,showpacs,preprintnumbers
,epsf,nofootinbib]{revtex4}
\usepackage{amsfonts,latexsym,eucal,amsmath,amssymb,times, ulem}
\usepackage[dvips]{graphicx}
\usepackage[usenames]{color}

\newcommand{\ie}{{\it i.e.}}
\newcommand{\eg}{{\it e.g.}}
\newcommand{\etal}{{\it et al.}}

\newcommand{\MBH}{{M_{\rm BH}}}
\newcommand{\MBHt}{{M_{\rm BH}^2}}

\newcommand{\mtot}{M}
\newcommand{\zeroth}{{[0]}}
\newcommand{\first}{{[1]}}
\newcommand{\Exp}{{\rm exp}}
\newcommand{\hybrid}{{\rm hyb}}
\newcommand{\xpole}{x_{\rm lr}}
\newcommand{\Rpole}{R_{\rm lr}}
\newcommand{\T}{{\rm T}}
\newcommand{\num}{{\rm full}}
\newcommand{\estimate}{{\rm guess}}

\definecolor{red  }{rgb}{1,0,0}
\definecolor{blue }{rgb}{0,0,1}
\definecolor{green}{rgb}{0,1,0}

\begin{document}

\thispagestyle{empty}

\title{Impact of the second-order self-forces on the dephasing 
of the gravitational waves from quasicircular extreme mass-ratio inspirals}
\date{\today}
\author{Soichiro Isoyama$^{1}$}
\email{isoyama_at_yukawa.kyoto-u.ac.jp}
\author{Ryuichi Fujita$^{2}$}
\email{ryuichi.fujita_at_uib.es}
\author{Norichika Sago$^{3}$}
\email{sago_at_artsci.kyushu-u.ac.jp}
\author{Hideyuki Tagoshi$^{4}$}
\email{tagoshi_at_vega.ess.sci.osaka-u.ac.jp}
\author{Takahiro Tanaka$^{1}$}
\email{tanaka_at_yukawa.kyoto-u.ac.jp}
\affiliation{\,\\ \,\\
$^{1}$ Yukawa Institute for Theoretical Physics, Kyoto university,
  Kyoto, 606-8502, Japan \\
$^{2}$ Departament de F\'isica, Universitat de les Illes Balears, 
 Palma de Mallorca, E-07122 Spain\\
$^{3}$ Faculty of Arts and Science, Kyushu University,
Fukuoka 819-0395, Japan\\
$^{4}$ Department of Earth and Space Science, Graduate School of Science, 
Osaka Univeristy, Osaka, 560-0043, Japan
 }

\preprint{YITP-12-66}


\begin{abstract}
The accurate calculation of the long-term phase evolution of 
gravitational wave (GW) forms from extreme (intermediate) 
mass-ratio inspirals [E(I)MRIs] is 
an inevitable step to extract information from this system. 
In order to achive this goal, it is believed that we need to understand 
the gravitational self-forces. 
However, it has not been quntatively demonstrated that  
the second-order self-forces are necessary for this purpose. 
In this paper we revisit the problem to estimate the order of magnitude 
of the dephasing caused by the second-order self-forces on a small body 
in a quasicircular 
orbit around a Kerr black hole, based on the 
knowledge of the post-Newtonian (PN) approximation and invoking the energy 
balance argument. 
In particular, we focus on the averaged dissipative part of the 
self-force, since it gives the leading-order contribution among 
their various components. 
To avoid the possibility of the energy flux of GWs becoming negative, 
we propose a new simple resummation called 
{\it exponential resummation}, which assures the positivity 
of the energy flux.
In order to estimate the magnitude of the 
yet-unknown second-order self-forces, here we point out 
the scaling property in the absolute value of the 
PN coefficients of the energy flux. 
Using these new tools, 
we evaluate the expected magnitude of dephasing. 
Our analysis indicates that the dephasing due to the second-order 
self-forces for quasicircular E(I)MRIs 
may be well captured by the 3 PN energy flux, 
once we obtain all the spin-dependent terms, 
except for the case with an extremely large spin of the 
central Kerr black hole. 
\end{abstract}


\maketitle


\section{Introduction}
\label{sec:Intro}

Extreme mass-ratio inspirals (EMRIs) 
and intermediate mass-ratio inspirals (IMRIs), in which a 
stellar mass or a compact object with several tens solar masses inspirals 
into a more massive central black hole, have attracted much interest 
not only as a promising source of the gravitational waves (GWs) for future 
spaceborne GW detectors, but also as a unique clean 
probe of the spacetime region of strong gravity. 
To achieve the test of general relativity using 
GWs from E(I)MRIs, we need to predict sufficiently accurate waveforms. 
This requirement motivates us to model the E(I)MRIs 
as the motion of a small body in a given background spacetime 
with gravitational backreaction.
{This backreaction is treated as  
gravitational self-forces
\cite{Mino:1996nk,Quinn:1996am,Galley:2006gs,Galley:2008ih,Gralla:2008fg,Pound:2009sm}, 
and its higher-order extension 
with respect to the mass-ratio has attracted much interest in recent years
\cite{Rosenthal:2005it,Detweiler:2011tt,Pound:2012nt,Gralla:2012db,Pound:2012dk}. 
See the following review articles, {\eg},~Refs.
\cite{Barack:2009ux,Poisson:2011nh} and references therein for more details.}

In fact, following the scaling argument
(See, {\eg}, Refs.\cite{Hinderer:2008dm,Tanaka:2005ue}) , 
the phase of GWs from a particle whose orbital 
frequency sweeps a few orders of magnitude before the plunge 
can be expanded as  
\begin{equation}\label{phase}
\Phi = \frac{\MBH}{\mu} 
\left[ \Phi^{(0)} + \frac{\mu}{\MBH} \Phi^{(1)} + 
O\left( \frac{\mu^2}{\MBHt} \right) \right], 
\end{equation}
where $\Phi^{(0)}$ and $\Phi^{(1)}$ are $O(1)$ quantities independent of
$\mu$, which is the mass of a small particle, and $\MBH$ and $a$ are 
the mass and the spin parameter of the Kerr black hole, respectively. 
\footnote{{Throughout this paper, we assume that the resonance is absent 
during its orbital evolution.
See Refs.\cite{Tanaka:2005ue,Mino:2005an,Flanagan:2010cd,Gair:2011mr,
Flanagan:2012kg} 
for further discussion about resonant orbits.}}
{On the one hand $\Phi^{(0)}$ in Eq.~\eqref{phase}, we only need 
the self-forces up to the first-order time-averaged dissipative part.
In addition, it has long been known that $\Phi^{(0)}$ can be 
computed with the well-established balance argument, which relates the 
first-order time-averaged dissipative part of the self-forces to the 
energy and angular momentum fluxes associated with global Killing vectors 
on the background Kerr black hole\cite{Gal'tsov:1982zz}.
Though the Carter constant is not associated with any Killing vector 
(and thus the simple balance argument is not applicable), 
the way to computing long time averaged evolution of the Carter constant 
has already been well established\cite{Mino:2003yg,Sago:2005gd,Sago:2005fn,Drasco:2005kz}. 
}
{On the other hand, $\Phi^{(1)}$, 
which is referred to as the dephasing relative to $\Phi^{(0)}$, 
depends on two different components of the self-forces: 
the first-order conservative part, 
and the averaged second-order dissipative part.
\footnote{Here the ``dissipative'' part refers to the self-forces 
that cause the time variation of the constants of motion, 
such as the energy,  
the angular momentum around the axis of symmetry, and the Carter
constant. The ``conservative'' part is the part that gives 
the correction to the relation between the orbital frequencies  
and the constants of motion\cite{Mino:2003yg,Tanaka:2005ue}.
The meaning of ``time-averaged'' is 
averaging over a sufficiently long period
compared to the time scale for the evolution of 
the phase difference between the oscillations in the 
radial and the zenith angle directions. }
The first subleading term, $\Phi^{(1)}$, can still be important, 
since potentially it may give a correction significantly 
greater than unity to the phase 
\cite{Burko:2002fd,Pound:2005fs,Pound:2007th,
Gair:2008bx,Thornburg:2010tq}. 

In the effort to obtain an accurate waveform, 
there have been many works on self-forces.
As for the first-order conservative part, 
thanks to recent massive development, 
we are now in part ready for practical computation with numerical implementation 
\cite{Barack:1999wf,Barack:2001gx,Mino:2001mq,Detweiler:2002mi,Barack:2002bt}. 
Particularly, in the case of the Schwarzschild background, 
the corrections to the orbital frequencies\cite{Warburton:2011fk}
and $\Phi^{(1)}$\cite{Lackeos:2012de} have already been studied extensively. 
Even in the case of the Kerr background  
preliminary results of the self-forces in quasicircular 
orbits have been reported\cite{Shah:2012gu}.  

By contrast, the averaged second-order dissipative part of the self-forces 
has so far been studied only at the formal level in the context of black hole
perturbation
\cite{Rosenthal:2005it,Detweiler:2011tt,Pound:2012nt,Gralla:2012db,Pound:2012dk}. 
It will require much more effort to establish the method for computing the 
second-order dissipative part, especially in the case of the Kerr background. 
Under such circumstances, a typical strategy for evaluating 
$\Phi^{(1)}$ is to make use of the standard post-Newtonian (PN) 
approximation, in which we assume slow motion of a satellite and its 
weak gravitational field.
Based on the PN approximation, Huerta and Gair\cite{Huerta:2008gb} 
evaluated the size of the dephasing caused by the first-order conservative 
self-forces and the averaged second-order self-forces, 
picking up representative EMRIs in quasicircular orbits on a Kerr black hole.
The same dephasing was also discussed by Yunes~{\etal}\cite{Yunes:2010zj}, 
using the effective one-body formalism again for representative E(I)MRIs.

A naive expectation is that the PN approximation will not suitable for 
modelling 
the waveforms of E(I)MRIs, especially for a Kerr black hole with large spin.
Typical E(I)MRIs in circular orbits spend 
the last few years of inspiral in the vicinity of the inner most stable 
circular orbit (ISCO). 
Since the ISCO radius reaches the event horizon as 
the spin of the Kerr black hole is increased to the extremal limit,
the motion of the body becomes highly relativistic, exceeding 
the validity range of the standard PN approximation\cite{Zhang:2011vh}.
However, things are not so trivial. The time spent near the 
ISCO becomes longer as we increase the mass ratio $\MBH/\mu$, but in that 
case the mass of the large central black hole also becomes larger. 
Then, the total cycles of GWs become smaller for a given observation time.
As a result, the correction due to higher-order self-forces might be
suppressed below the observational threshold, 
despite the loss of accuracy of the PN approximation.
Therefore, it is not so obvious whether there are really E(I)MRIs that 
require the notion of the second-order self-forces. 

The previous analyses mentioned above\cite{Huerta:2008gb,Yunes:2010zj} 
are focusing on the corrections 
coming from the self-forces at the currently available PN order 
and are limited to representative E(I)MRIs. 
To get an insight into whether or not the second-order self-forces based 
on the black hole perturbation 
are really necessary to calculate the waveforms of 
quasicircular E(I)MRIs, therefore, 
it would be useful to give an estimate of dephasing coming 
from the averaged second-order dissipative self-forces, 
focusing on the yet-unknown higher PN terms and  
surveying the whole parameter region of E(I)MRIs.

What we discuss in this paper is the adiabatic evolution of E(I)MRIs 
in quasicircular orbits on a Kerr spacetime. 
Here the adiabatic evolution means an approximation in which 
the evolution of the orbital frequency is determined by the energy balance 
argument; \ie, the rate of change of the total energy of the binary 
is equated to the energy flux emitted to infinity.  

To evaluate the order of magnitude of the yet-unknown higher PN
corrections, we need to rely on some extrapolation. 
For this purpose, we first introduce a simple new resummation 
of the energy flux, which we call the ``exponential resummation.''  
When the spin of a Kerr black hole is large enough, 
the PN energy flux in the Taylor form can 
be negative outside the ISCO radius for some PN orders\cite{Tagoshi:1996gh}. 
If this happens, the estimated total phase before the plunge diverges 
and the extrapolation to the higher PN order will not make sense. 
Our simple ``exponential resummation'' is the one that ensures the positivity 
of the energy flux. 

As a PN input for the corrections at the next leading order in the mass ratio, 
\footnote{
Throughout this paper, 
we use the notion ``-mass ratio'' to refer to the symmetric mass ratio 
$\nu := \mu\MBH/ (\MBH + \mu)^2$, instead of the usual mass ratio 
$\mu/\MBH$. 
We should note that the terms higher-order in the mass ratio arise 
even at the level of the quadrupole formula if we use another mass ratio. 
In this sense, definitely the use of the symmetric mass ratio is advantageous. 
This fact has indeed been taken into account in the previous 
analyses such as that of Le Tiec~{\etal} \cite{LeTiec:2011bk}, 
where the periastron 
advance due to the conservative portion of the first-order self-forces 
is compared to the result deduced from the numerical relativity.
}}
the best one available so far 
is the 3.5PN energy flux of GWs\cite{Blanchet:2001aw,Blanchet:2001ax}  
with linear spin-dependent terms up to the 3 PN order, 
which has recently been derived by 
Blanchet {\etal}\cite{Blanchet:2011zv}. 
To estimate the possible magnitude of the yet-unknown 
higher-order PN terms, we focus on a scaling property 
among the PN coefficients in the energy flux.  
Using the 8 PN energy flux in the test particle limit
\cite{Fujita:2011zk, Fujita2012:un}, 
we will show that the absolute values of the coefficients 
scale roughly as required from the convergence of the PN series 
up to the light-ring radius. 
(This point was also discussed by Nakano~{\etal} 
independently\cite{Nakano2012:pc}.)
Since this scaling behavior is related to the PN convergence, 
despite the lack of the higher PN terms in the energy flux, 
we conjecture that the same scaling property will hold 
for the higher-order terms in the mass-ratio. 
Under this assumption, we estimate 
the order of magnitude of the unknown portion of the energy flux 
coming from higher PN terms at the next leading order in 
the mass ratio via ``extrapolation.'' 
Gathering the tools stated above, we will investigate the 
impact on the dephasing from the averaged dissipative second-order 
self-forces for various E(I)MRIs systematically. 
This is the main goal of this paper.

The remainder of this manuscript is organized as follows:
In Sec.~\ref{sec:N}, 
we briefly review how the accumulated phase of GWs
from adiabatic inspiral is calculated based on the balance argument. 
In Sec.~\ref{sec:exp-resum}, 
to cure the negative energy flux of GWs that appears 
in the truncated PN Taylor series expansion, 
we propose the exponential resummation that ensures 
the positivity of the energy flux. 
Section~\ref{sec:8PN} is dedicated to the study of the scaling property 
in the coefficients of the energy flux in the test particle limit
that becomes manifest owing to the brand-new 8 PN energy flux
\cite{Fujita2012:un}.
In Sec.~\ref{sec:N-PN}, 
using the exponential resummation and the scaling property, 
we will estimate the dephasing coming from the yet-unknown part of the 
second-order self-forces for various E(I)MRIs.
We find that the unkown non-linear spin-dependent terms 
at the lower PN order in the energy flux dominate the unknown dephasing.  
We summarize our results and conclude in Sec.~\ref{sec:summary}.


In this manuscript, we use geometrical units $G = c = 1$, and 
the sign convention of the metric is $(-,+,+,+)$. 
The coordinates $(t, r, \theta, \phi)$ denote the Boyer-Lindquist 
coordinates of the Kerr black hole\cite{Boyer:1966qh}.
We frequently use the dimensionless spin defined by $q := a / \MBH$, 
and the symmetric mass ratio defined by $\nu := \mu\MBH/ (\MBH + \mu)^2$.

\section{The accumulated phase of the gravitational wave from an inspiraling 
binary}
\label{sec:N}
We consider a binary composed of a small satellite body with 
the rest mass $\mu$ in a quasicircular orbit around a 
Kerr black hole with the mass $\MBH$ and the spin parameter $a$.
We neglect the effect of the spin of the satellite, which may be negligiblly 
small for the detection stage of E(I)MRIs, though it plays no negligible role 
in their parameter estimation \cite{Huerta:2011kt,Huerta:2011zi}.
For a binary in a quasicircular orbit, 
the accumulated phase of GWs is calculated as 
\begin{eqnarray}
\label{def-cycle}
\Phi &:=& -2 \int_{x_{\rm{ISCO}}}^{x_0} dx \frac{x^{3/2}}{\mtot}
\frac{E' (x) }{  \dot{E}(x) },
\end{eqnarray}
where $E$ is the binding energy of the binary, 
$\dot{E} :=dE /dt$ is the energy loss rate, 
and the prime denotes the differentiation with respect to $x$, 
which is a dimensionless orbital 
frequency defined by $x := (\mtot \Omega)^{2/3}$, 
with the orbital frequency $\Omega$ and 
the total mass $\mtot := \MBH + \mu$. 
Here $E$ is supposed to depend only on $x$, neglecting the effect of the 
variation of mass due to the energy absorbed by the black hole. 
Throughout the paper we totally neglect the effect of 
this time-dependent mass variation.
We choose the lower bound of the integral in Eq.~\eqref{def-cycle}, 
$x_{\rm{ISCO}}$, to be the value of $x$ 
at the inner most stable circular orbit (ISCO), 
and the upper bound, $x_0$, to be the value determined by the 
condition coming from the finite observation time.
Throughout this paper, we adopt 
$t_{\rm{obs}} = 1~{\rm{yr}}$ as the observation time. 

Our primary interest in this paper is the order of magnitude of the dephasing 
coming from the corrections at the next leading order in the mass ratio. 
To evaluate it, we calculate the difference 
between $\Phi$ with and without the higher-order corrections. 
In checking the detectability of this difference by observation, 
the initial and the final frequencies should be kept unchanged 
when we evaluate it. 
Here, we fix both $x_0$ and $x_{\rm{ISCO}}$ to the values 
determined by the test particle limit. 
Namely, \cite{Bardeen:1972fi}
\begin{eqnarray}\label{x-ISCO}
x_{\rm{ISCO}} &:=& \left( {  R_{\rm{ISCO}}^{3/2} + q  } \right)^{-1/6},
\qquad
R_{\rm{ISCO}} :=  3 + Z_2 \mp \sqrt{(3 - Z_1) (3 + Z_1 + 2 Z_2)}, \cr
Z_1 &:=& 1 + (1 - q^2)^{1/3} [(1 + q)^{1/3} + (1 - q)^{1/3} ],
\quad
Z_2 := \sqrt{3 q^2 + Z_1^2},
\end{eqnarray}
where the upper (lower) sign in the second equation 
is chosen for the co- (counter-)rotating case with $q > (<) 0$, 
and $R := r / \MBH$ represents the dimensionless radius. 
$x_0$ is determined by 
\begin{equation}
\label{test-year}
1~\mbox{(yr)} = - \int_{x_{\rm{ISCO}}}^{x_0} 
{dx} \frac{ E'{}^{\zeroth} (x)}{ \dot{E}^{\zeroth} (x) },
\end{equation}
where the superscript $^{\zeroth}$ means the leading-order contribution in 
the mass ratio, \ie, the test particle limit, and 
\begin{eqnarray}\label{circular-EL}
E^{\zeroth} = \mu \frac{R^{3/2} - 2 R^{1/2} + q }
{ R^{3 / 4} (R^{3/2} - 3 R^{1/2} + 2 q )^{1/2} }, 
\qquad
\Omega := \frac{1} {M ( R^{3/2} + q )},
\qquad
R = \left( \frac{1 - q x^{3/2}}{ x^{3/2} } \right)^{2/3}.
\end{eqnarray}

As for the energy, we adopt here the binding energy of the binary instead 
of the energy directly related to the four-momentum of the particle. 
It would be more natural to consider the latter 
in the context of the self-force calculation based on 
the black hole perturbation, but it is not gauge invariant. 
Hence, the former is more suitable for comparison with the 
PN calculation. These two energies are different at the next leading order 
in the mass ratio due to the presence of the gravitational field 
energy, but there must be one-to-one correspondence between them, 
once the gauge is completely fixed.

The energy balance argument tells us that the averaged loss of the 
total energy should be equal to the averaged total energy flux to the  
future null infinity because of energy conservation 
(see,~{\eg}, Refs.\cite{Geroch:1981ut,Gal'tsov:1982zz}). 
Hence, as long as we define the binding energy appropriately, 
\begin{equation}\label{balance}
- \frac{dE}{dt} = {\cal{L}}
\end{equation}
holds, in the sense averaged over a sufficiently long time 
and neglecting the horizon absorption flux,
where ${\cal{L}}$ is the energy flux emitted to the null infinity. 
Hence, we evaluate the phase by  
\begin{equation}\label{phase-evo}
\Phi[{\cal L}]:= 
2 \int_{x_{\rm{ISCO}}}^{x_0} dx 
\frac{x^{3/2}}{\mtot} \frac{E'{}^{\zeroth}(x)}{ {\cal L} (x)}. 
\end{equation}
In the above equation, 
as we are interested in the dissipative corrections, we 
fix $E'$ to the expression in the test particle limit. 
We expand ${\cal L}$ as 
${\cal L} = {\cal L}^{\zeroth}  + \nu {\cal L}^{\first} + O(\nu^2)$. 
To compute ${\cal L}^{\zeroth}$, we can use
the Teukolsky formalism\cite{Teukolsky:1973ha} 
and invoke the numerical code developed by 
Fujita and Tagoshi\cite{Fujita:2004rb, Fujita:2009uz}. 

The finite mass corrections at the next leading order, ${\cal L}^{\first}$,
are in part provided by the standard PN calculations. 
In the PN formalism, the energy flux of GWs emitted to infinity 
from a quasicircular binary 
is obtained up to the 3.5PN order for the spin-independent 
terms\cite{Blanchet:2001aw,Blanchet:2001ax} 
and up to the 3 PN order for the terms linear in spin
when the spin vectors are parallel to the orbital 
axis\cite{Blanchet:2011zv}.
In the present notation, truncated at $O(\nu)$, it is given by 
\begin{eqnarray}\label{3PN-flux}
{\cal L}^\T_{n {\rm PN}} (x,q) &=&
{32 \over 5} x^5 \nu^2 \biggl{[}
1 + x \left( -\frac{1247}{336} - \frac{35}{12}\nu \right)
+ x^{3/2} \left( 4 \pi - {11\over 4} q + {17 \over 4} q\nu \right) 
+ x^2 \left( -{44711 \over 9072} + {9271 \over 504} \nu \right) \cr
&&+ x^{5/2} \left( - {8191 \over 672} \pi - { {59} \over 16} q 
+ \nu \left\{ 
- {583 \over 24}\pi + {3749 \over 144} q \right\} \right) \cr
&&+ x^3 \biggl( {6643739519 \over 69854400} + {16 \over 3} \pi^2 
- {1712 \over 105} \gamma_{\rm E} - {856 \over 105} \log (16 x) - 
  {65 \over 6} \pi q 
+ \left\{ 
- {134543 \over 7776} + {41 \over 48} \pi^2 +{33\over 2}\pi q \right\} \nu
  \biggr) 
 \cr & &+ 
 x^{7/2}\left(-{16285\over 504}+{214745\over 1728}\nu\right)\pi
 +O(x^4,\nu^2,q^2,q x^{7/2})
 \biggl]~, 
\end{eqnarray}
where $\gamma_{\rm E} = 0.57721 \dots$ denotes 
Euler's constant.
We will refer to the above expression for 
the energy flux truncated at the $n$th PN order
\footnote{
We refer to terms of $O(x^{n})$ relative to the 
leading order as those of $n$PN order.
}
as ``the $n$PN Taylor flux.'' We 
denote it by ${\cal L}^\T_{n {\rm PN}}$, and expand it like 
${\cal L}^\T_{n \rm PN}
:={\cal L}^{\T\zeroth}_{n \rm PN} + 
\nu {\cal L}^{\T\first}_{n \rm PN}+\cdots$ as before. 
We also frequently use the ``normalized $n$PN flux'' defined by 
\begin{eqnarray}
{L}^{\T}_{n \rm PN}(x,q)
:= \left(\frac{32}{5} \nu^2 x^5 \right)^{-1} {\cal L}^{\T}_{n \rm PN}(x,q).
\end{eqnarray}
From now on, we denote the PN flux solely composed of the known part 
 as ``the known $n$PN Taylor flux'' and distinguish 
it from the full PN flux with the notation $\tilde{}$, as with 
${\cal \tilde L}^{\T}_{n \rm PN}$. 
We also introduce notation to denote 
the residual terms of higher-order in PN expansion and in spin: 
\begin{equation}
{\cal L}^{\T [i]}_{>n \rm PN}
:= {\cal L}^{[i]}_{\rm full} - {\cal L}^{\T [i]}_{n \rm PN}. 
\end{equation}
Then, ${\cal \tilde L}^{\T\first}_{>3.5 \rm PN}$
represents the yet-unknown energy flux to be determined 
from the computation of the second-order dissipative self-forces. 
Further we define 
${\cal \tilde L}^{\T\zeroth}_{n\rm PN}$ and 
${\cal \tilde L}^{\T\zeroth}_{>n\rm PN}$ 
by the terms in ${\cal L}^{\zeroth}$ that correspond to 
${\cal \tilde L}^{\T\first}_{n\rm PN}$ and 
${\cal \tilde L}^{\T\first}_{>n\rm PN}$, respectively.  
To be precise, we define ${\cal \tilde L}^{\T\zeroth}_{n\rm PN}$ 
by the sum of the spin-independent terms up to the 3.5PN order 
and the terms linear in the black hole spin up to the 3PN order. 
${\cal \tilde L}^{\T\zeroth}_{>n\rm PN}$ signifies the remaining terms 
at leading order in the mass ratio, 
which also includes the nonlinear spin-dependent terms in all PN orders.
\footnote{Here we do not distinguish 
the logarithmic term in the 3 PN Taylor flux in the test particle limit 
from the other 3 PN terms. 
We have confirmed that the results in the present paper do not 
change much even if we exclude this term from 
${\cal \tilde L}^{\T\zeroth}_{n\rm PN}$}.

\section{A manifestly positive definite energy flux: exponential 
resummation }
\label{sec:exp-resum}

In this section we propose a new simple resummation scheme, which 
can be easily applied if we just know the $n$PN Taylor flux. 

\subsection{Limitation of the Taylor flux revisited}
\label{subsec:dEdt}
It is easy to imagine that the PN Taylor flux is not so accurate 
when the orbital radius becomes small. 
Indeed, Zhang~{\etal}\cite{Zhang:2011vh} pointed out that 
the $n$PN Taylor flux in the test particle limit 
rapidly loses accuracy around the ISCO radius.  
However, it is a different issue whether the Taylor flux is accurate 
enough for our present purpose 
because what we are interested in here is whether or not 
the dephasing in GW waveforms is measurbale.

As we want to evaluate the correction to $\Phi$ in Eq.~\eqref{phase-evo}
caused by the yet-unknown part of the flux 
${\cal \tilde L}^{\T \first}_{>n \rm PN}$, 
we need some extrapolation method. 
For this purpose, it will later become necessary to evaluate something 
like the phase for the flux truncated at the $n$PN order, 
$\Phi[{\cal \tilde L}^{\T \zeroth}_{n \rm PN}]$. 
At this point, we find that the $n$PN Taylor flux is problematic. 
As mentioned earlier, 
the energy flux truncated at some PN orders 
becomes negative outside $R_{\rm ISCO}$ 
if the spin of black hole $q$ is sufficiently  
large, firstly pointed out by~Tagoshi~{\etal}\cite{Tagoshi:1996gh}.

\begin{figure}[tbp]
\hspace{-3cm}
   \begin{center}
    \includegraphics[width=8cm, clip, angle = -90]
    {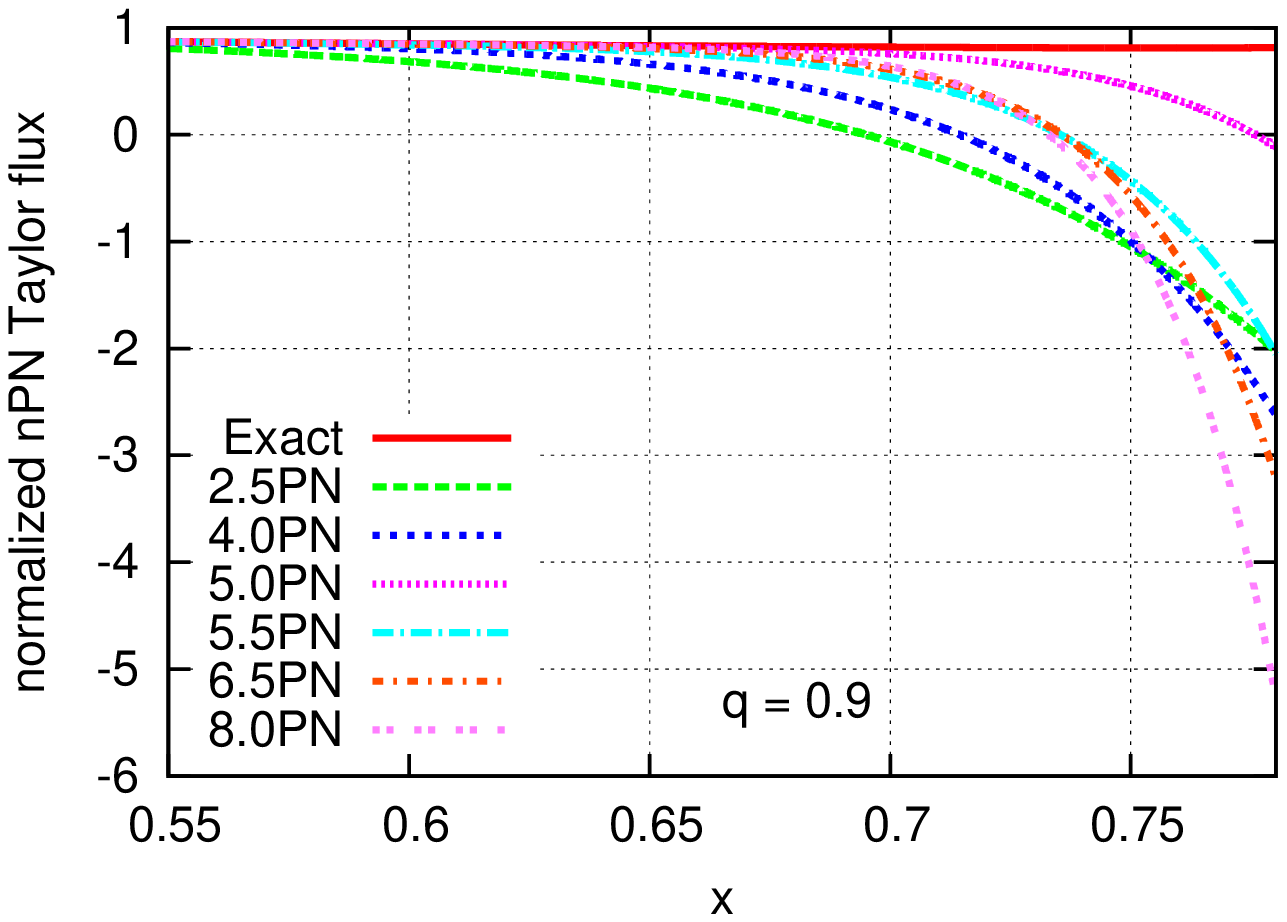}
  \end{center}
 \vskip -\lastskip \vskip -3pt
   \caption{
    The normalized $n$PN Taylor flux in the test particle limit, 
    ${L}^{\T\zeroth}_{n \rm PN}(x,q)$,
    for $q = 0.9$ up to $x_{\rm ISCO}(0.9) :=0.78014\dots$. 
   The horizontal axis is the dimensionless frequency $x$ of 
    a body. 
   The label ``Exact'' means the exact numerical energy flux, 
   calculated by Fujita and Tagoshi\cite{Fujita:2004rb}.
  }
  \label{fig:dEdt-PN}
\end{figure}

To see how serious the problem is,  
we revisit the Taylor flux in the test particle limit.
The spin-independent terms in 
${\cal L}^{\T\zeroth}_{n \rm PN}$ 
have been analytically calculated up to the 22 PN order 
and the spin-dependent terms up to the 8 PN order 
by Fujita\cite{Fujita:2011zk, Fujita2012:un}.
The expression of the normalized 8 PN energy flux including complete 
spin-dependent terms is schematically expanded as 
\footnote{
The explicit expression for ${\cal L}^{\T\zeroth}_{8\rm PN}$ 
will be made available
elsewhere in an appropriate form\cite{Fujita2012:un}.
}
\begin{eqnarray}\label{Kerr-8PN}
L^{\T\zeroth}_{8 \rm PN}(x,q)
= 
\sum_{n=0}^{8} \sum_{p=0}^{p_{\rm max}}
\hat L^{(n,\,p)}(q)  x^n (\log(x))^p, 
\end{eqnarray}
where $p_{\rm max}$ is the maximum integer that does not exceed $n/3$.
In Fig.~\ref{fig:dEdt-PN}, 
we depict $L^{\T\zeroth}_{n \rm PN}(x,q)$, 
the normalized $n$PN Taylor flux  
in the test particle limit truncated at various PN orders for $q = 0.9$.
The curves terminate at the ISCO frequency. 
We also plot the exact numerical energy flux 
in the test particle limit as a reference.
The exact numerical energy flux is manifestly positive definite 
for $x<x_{\rm ISCO}(0.9)$, while all the $n$PN Taylor fluxes 
in Fig.~\ref{fig:dEdt-PN} 
cross zero at some $x = x_0(q) < x_{\rm ISCO}(0.9)$.  
In addition to the already-known 2.5 PN and 4 PN cases 
\footnote{While Tagoshi~{\etal}\cite{Tagoshi:1996gh} reported 
that the 3 PN Taylor flux with $q = 0.9$ became negative for small radii, 
we find that the 3 PN Taylor flux is always positive definite for 
all radii, irrespective of the spin parameter of the Kerr black hole.}
\cite{Tagoshi:1996gh}, we also find that the 5 PN, 5.5PN, 6.5PN,
and 8 PN fluxes become negative before $x$ reaches $x_{\rm ISCO}(q)$. 
Even for a moderate value such as $q=0.7$,
we still observe the flux to cross zero, say at order 8PN. 
Once this happens, the integrand of~\eqref{phase-evo} diverges, 
and then the phase evaluated by using the truncated flux 
${\cal L}^{\T\zeroth}_{n \rm PN}$ does not make sense. 
For a reliable extrapolation, it is necessary to use a resummed 
expression for the energy flux that gives at least 
a finite estimate of the phase for the truncation 
at any PN order.


\subsection{The exponential resummation and an improved hybrid energy flux}
\label{subsec:Hybrid}
To overcome the drawback of the $n$PN Taylor flux, 
a naive requirement will be to guarantee that the energy flux 
is always positive by resummation.
Among various known resummation 
techniques\cite{Damour:1997ub,Porter:2004eb,Porter:2007vk,Carre:2012wm}, 
to the best of our knowledge, 
only the factorized resummation ensures the positivity of the energy
flux. This resummation was proposed by Damour {\etal}\cite{Damour:2008gu} 
for the equal-mass nonspinning binaries in a circular orbit. 
Fujita and Iyer\cite{Fujita:2010xj} and Pan {\etal}\cite{Pan:2010hz} 
applied it to a test particle in a circular orbit around 
a Schwarzschild and a Kerr black hole, respectively.
Here, we propose another even simpler resummation, 
which is defined by
\begin{eqnarray}
\label{exp-resum}
{\cal L}^{\Exp}_{n \rm PN} (x, q) := \frac{32}{5} \nu^2 x^5 
\exp[L^{\Exp}_{n \rm PN}(x, q)],
\end{eqnarray}
with 
\begin{equation}
L^{\Exp}_{n \rm PN}(x, q):=\log\left[
 {L}^\T_{n' \rm PN}(x,q)\right]
 \Bigr\vert_{{\rm truncated~ at~}n\rm PN~order}, 
\end{equation}
where $n'\geq n$ is understood. 
We refer to this resummation and the energy flux [Eq.~\eqref{exp-resum}] 
as the ``exponential resummation'' and  
the `` $n$PN exponential resummed flux,'' respectively.  
This flux is manifestly positive 
and can incorporate any higher-order corrections 
if the counterparts in the $n$PN Taylor flux are given. 
{Furthermore, it significantly improves the accuracy of approximation
\cite{Fujita2012:un}.
For definiteness, we explicitly write the definitions of 
${\cal L}^{\Exp\zeroth}_{n \rm PN}$ and ${\cal L}^{\Exp\first}_{n \rm PN}$ 
as 
\begin{equation}
{\cal L}^{\Exp\zeroth}_{n \rm PN} (x,q):= \frac{32}{5} \nu^2 x^5 
\exp[L^{\Exp\zeroth}_{n \rm PN}(x, q)],
\qquad 
{\cal L}^{\Exp\first}_{n \rm PN} (x,q):= 
\left( 
\frac{32}{5} \nu^2 x^5 
\exp[L^{\Exp\zeroth}_{n \rm PN}(x, q)] \right)
L^{\Exp\first}_{n \rm PN}(x, q) ,
\end{equation}
with $L^{\Exp}_{n\rm PN}=L_{n\rm PN}^{\Exp\zeroth}
+\nu L_{n\rm PN}^{\Exp\first}+\cdots$. 
We show the explicit form only for $L^{\Exp\first}_{n \rm PN}$ as  
\begin{eqnarray}
 L^{\Exp\first}_{n \rm PN}(x,q)
 &=& 
 - {35 \over 12} x + {17 \over 4} q x^{3/2} + {30523 \over 4032} x^2 
+ \left( \frac{136229}{4032} q - \frac{101}{8}\pi \right) x^{5/2} \cr
&& \quad
+ \left(  - {43670915 \over 12192768} + \frac{41}{48}\pi^2 
-{\pi \over 2} q  \right) x^3 
+ \frac{70075}{6048} \pi  x^{7/2} +O(x^4,q x^{7/2},q^2)~.
\end{eqnarray}
}

Combining the exact numerical flux obtained 
in the test particle limit, ${\cal L}^{\zeroth}_{\num}$, 
and the PN flux, ${\cal L}_{\rm PN}$, 
one can obtain a better estimate for the energy flux. 
In the case of the Taylor flux, these two are 
combined simply by summing ${\cal L}^{\zeroth}_{\num}$ and 
${\cal L}^{\T\first}_{\rm PN}$.  
In the case of the exponential resummed energy flux [Eq.~\eqref{exp-resum}], 
we need to do this summation at the level of the exponent, $L^\Exp$, say, 
expanding $L_{n\rm PN}^\Exp$ in powers of $\nu$, 
we replace $L_{n\rm PN}^{\Exp\zeroth}$ with the one corresponding 
to the numerical flux. 
Then, we obtain
\begin{eqnarray}
\label{hybrid}
{\cal L}^\hybrid_{n\rm PN}(x,q)
= {\cal L}^{\zeroth}_{\num}(x,q) 
\exp \left[ \nu L^{\Exp\first}_{n \rm PN}(x,q) +\cdots \right], 
\end{eqnarray}
which we call the ``$n$PN hybrid flux.'' 

Analogously, we introduce the ``known $n$PN exponential resummed flux'' and 
the ``known $n$PN hybrid flux,'' and distinguish them from their respective
counterparts with the notation $\tilde{}$.  
Here, the truncation for the known part 
is made at the level of ${L}^{\Exp\first}$.
Namely, ${\tilde L}^{\Exp\first}_{n\rm PN}$ is truncated 
at the same PN order for each 
order of spin that is included in the known terms 
${\cal \tilde L}^{\T\first}_{n\rm PN}$. 
For the exponential resummed flux, we can also introduce 
${\cal \tilde L}^{\Exp\zeroth}_{n\rm PN}$, 
the counterpart of ${\cal \tilde L}^{\Exp\first}_{n\rm PN}$ in 
the test particle limit, as in the case of 
${\cal \tilde L}^{\T\zeroth}_{n\rm PN}$. 
That is, ${\tilde L}^{\Exp\zeroth}_{n\rm PN}$ is truncated 
at the same PN order for each 
order of spin that is included in the known terms 
${\cal \tilde L}^{\T\first}_{n\rm PN}$. 
On the other hand, 
${\cal \tilde L}^{\hybrid \zeroth}_{n\rm PN}$ 
does not make sense. 
In the following discussion, we identify the hybrid flux with the 
exponential resummed flux for the leading order in the mass ratio. 
We summarize our notation for various fluxes 
used in the rest of our paper in Table~\ref{table:flux}, for readability.

\begin{table}[htb] 
\begin{tabular}{c|c}
\hline\hline
\mbox{}  & Symbol  \\
\hline\hline
The flux truncated at $n$PN order & ${\cal L}_{n\rm PN}^{[i]}$ \\
The residual part of ${\cal L}_{n\rm PN}^{[i]}$ 
&  ${\cal L}_{>n \rm PN}^{[i]}$ \\
The known part of the flux truncated at $n$PN order 
& ${\cal \tilde L}_{n\rm PN}^{[i]}$ \\
The residual of ${\cal \tilde L}_{n\rm PN}^{[i]}$ 
& ${\cal \tilde L}_{>n \rm PN}^{[i]}$  \\
The normalized $n$PN Taylor flux 
& $L^{{\rm T} [i]}_{n\rm PN}$ \\
The exponent of $n$PN exponential resummed flux
& $L^{{\rm exp} [i]}_{n\rm PN}$ \\
\hline\hline
\end{tabular}
\caption{Our notation for various fluxes in this paper. Upper index ${[i]}$ 
refers to the order of truncation with respect to the mass ratio $\nu$: 
${\cal L} = {\cal L}^{\zeroth}  + \nu {\cal L}^{\first} + O(\nu^2)$.
We also put label ``T'', ``exp'', ``'hyb'' and ``full'' to distinguish 
the flux type. Respectively, they correspond to 
Taylor, exponential, hybrid and numerical complete flux.}
\label{table:flux}
\end{table}

\section{The scaling law of the coefficients in the energy flux}
\label{sec:8PN}
Now we tackle our main issue: 
how to evaluate the magnitude of the yet-unknown 
part of the energy flux, ${\cal \tilde L}^{\first}_{>3.5 {\rm PN} }$.
(When we discuss general issues independent of the 
form of the energy flux, we simply suppress the labels 
``$\T$'', ``$\Exp$'' and ``$\hybrid$.'') 
Our strategy here is to establish the scaling property 
in the coefficients of the PN expansion of the energy flux in 
the test particle limit for fixed $q$. 
(This point was also discussed by Nakano~{\etal} 
independently\cite{Nakano2012:pc}.)
The argument of the $n$PN exponential resummed flux, 
${L}^{\Exp\zeroth}_{n \rm PN}$, is expanded as 
\begin{eqnarray}
\label{14PN}
L^{\Exp\zeroth}_{n \rm PN}(x,q)
:= 
\sum_{n}\left({x\over \xpole(q)}\right)^n 
\sum_{p} C_{n,p}(q)  
\left(\log\left[{x\over \xpole(q)}\right]\right)^p,
\end{eqnarray}
with 
\begin{equation}\label{light-ring}
\xpole (q) := \left( \frac{1}{ \Rpole^{3/2} + q } \right)^{2/3},
\end{equation}
where 
\begin{equation}
\Rpole (q) := 
2 \left[ 1 + \cos \left(
\frac{2}{3} \arccos q \right)  \right] 
\end{equation}
is the value of $x$ corresponding to the radius of the circular orbit 
on the light ring\cite{Bardeen:1972fi}. 



\begin{figure}[tbp]
\begin{center}  
\hspace{-3cm}
    \qquad \qquad 
    \includegraphics[width=8cm, clip, angle = -90]
    {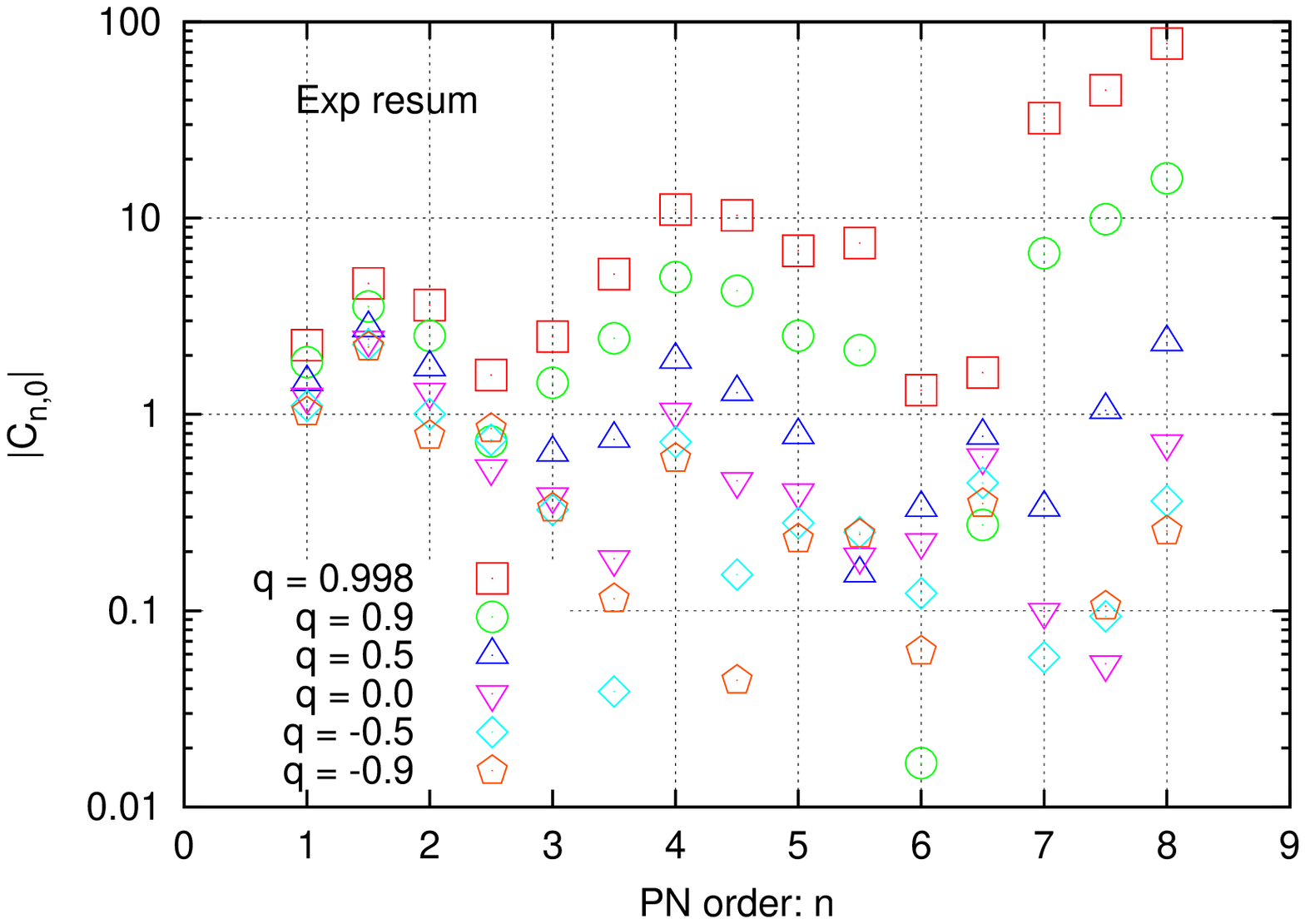}
 \end{center}
 \vskip -\lastskip \vskip -3pt
   \caption{
     The absolute value of the contribution to the energy flux 
      from each PN order $|C_{n,0}|$ at $x=\xpole(q)$,  
      with the horizontal axis being the PN order $n$. 
      Note that the values of $|C_{n,0}|$ are accidentally small 
      with some fixed $q$. We have safely exludeded these values from the plot.
}
  \label{fig:scaling}
\end{figure}%
It is known that for circular orbits the source term of the Teukolsky equation 
has a simple pole when the orbit is on the light-ring radius, \ie, at 
$x=\xpole(q)$\cite{Cutler:1993vq,Sasaki:2003xr}. 
The light-ring radius is the innermost radius for the presence of 
an (unstable) circular orbit, and there the particle energy per unit rest mass 
diverges. 
As a result, the energy flux also diverges there. 
Besides the light-ring singularity, another origin of singularity in solving 
the Teukolsky equation is the presence of quasi-normal mode poles, 
which are zeros of the Wronskian. 
However, the absolute value of the first quasi-normal mode frequency, 
which is the smallest, is almost identical to the frequency of GWs 
from the source at the light ring\cite{Onozawa:1996ux}. 
Hence, we expect that 
$\xpole$ determines, at least approximately, the convergence radius 
when the energy flux is seen as an analytic function of $x$.  
If the PN expansion of the energy flux converges 
as long as $x < \xpole(q)$, then $x=\xpole(q)$ is the first place 
where the series ceases to converge. 

With this expectation in mind, 
we plot the absolute value contribution from each 
PN order $|C_{n,0}|$ with $x=\xpole(q)$ 
in Fig.~\ref{fig:scaling} for various values of $q$. 
Here $q = 0.998$ is a bound for possible maximum spin of an 
astrophysical black hole,  called the Thorne limit\cite{Thorne:1974ve}.
As expected, $|C_{n,0}|$ is roughly independent of the PN order $n$, 
as is typical at the boundary of convergence for $|q| < 0.9$.  
Contrary to our expectation, for $|q| > 0.9$, 
$|C_{n,0}|$ continues to increase up to the 8 PN, 
and the series seems to diverge. 
However, this does not mean that PN series does not converge 
for $x\approx \xpole(q)$ with $|q|\approx 1$. 
When we make a similar plot for each partial wave contribution 
for a higher multipole component, the contribution to the energy flux 
peaks around a higher PN order, but the series is convergent,  
although we cannot check extremely high multipole components. 
On the other hand, the summation over multipole components 
is known to converge from the numerical computation of the 
energy flux 
by Fujita and Tagoshi\cite{Fujita:2004rb, Fujita:2009uz}. 
These facts may indicate that the convergence depends on the order of 
summation. 
The point of Fig.~\ref{fig:scaling} is that  $|C_{n,0}|$ 
shows a nice scaling property with respect to the PN order 
for any value of $q$. 
Even for $q$ close to the extremal limit, the plot becomes 
flat if we substitute a slightly smaller value of $x$ instead 
of $\xpole$. 
 
Now we are in the position to discuss how to estimate the order of magnitude 
of the yet-unknown part of the energy flux, 
${\cal \tilde L}^{\first}_{> 3.5 {\rm PN}}$.
The second-order perturbed Einstein equation would schematically take 
the form 
$\square h^{(2)} =  (T[z + \delta z] - T[z]) + h^{(1)} T[z] 
+ (\nabla h^{(1)})^2$, 
where $h^{(1)}$ and $h^{(2)}$ are, respectively, the first- and second-order 
perturbations induced by a body. 
Here $z$ is the background Kerr geodesic and $\delta z$ is the 
$O(\nu)$ correction to it\cite{Poisson:2004gg}.
We know that all the first-order perturbations have singularity 
only at $x\approx \xpole(q)$. 
The Green's function $\square^{-1}$ 
is basically the same as in the linear case. 
Hence, even for the second-order perturbation, the convergence radius 
in $x$ will be the same as in the linear case, and thus a similar scaling 
for the PN coefficients will be expected for the next leading order in 
the mass ratio, too. Then, we can guess the amplitude of the 
higher PN coefficients from the first few terms in the PN expansion 
that are known from the standard PN calculation. 

One may worry that $\square^{-1}$ has singularities at 
the quasi-normal mode frequencies. When we consider the second-order 
perturbation, there may arise high-frequency components which are absent 
at the level of linear perturbation. However, this might not happen.  
The point is that the metric perturbation caused by 
a circular geodesic has a helical Killing vector.
\footnote{
	The helical Killing vector $t^{\alpha} + \Omega \phi^{\alpha}$, 
	that remains to generate a symmetry even for the perturbed spacetime: 
	where $t^{\alpha}$ and 
$\phi^{\alpha}$ are the asymptotically time translation 
and rotational Killing vectors of the background Kerr black hole, 
respectively. See,~{\eg}, Ref.\cite{Friedman:2001pf} 
for more detailes. }
At the level of the second-order source term, 
this helical symmetry is broken due to the presence of the 
deviation from the geodesic, $\delta z$. 
Apart from this contribution, however, 
the source term keeps the helical symmetry and hence the frequency 
that appears in a partial wave labeled by 
$(\ell,m)$ is $m\Omega$ only. Namely, higher frequency modes do not arise. 
Even if we take into account $\delta z$, the time scale associated with 
$\delta z$ is as slow as $\Omega^{-1} \times O(\nu^{-1})$. 
Therefore, the presence of $\delta z$ alters the frequency $m\Omega$ 
only by a small amount of $O(\nu \Omega)$. 
This essentially does not change the convergence radius in $x$.

Even if we find the order of magnitude of the PN coefficients, 
the scaling property tells us nothing about the actual sign of each term. 
Therefore, when $x$ is close to $\xpole(q)$, 
it becomes difficult to guess the order of the 
magnitude of the infinite summation.  
However, since the higher-order PN terms are suppressed by the 
power of $x/\xpole(q)$, the summation is dominated by a few 
leading terms when $x/\xpole(q)$ is reasonably small. 
In that case, we do not have to worry about the infinite summation. 
Here, we bravely step a little forward by making the following proposal: 
the ratio between 
${\cal \tilde L}^{\first}_{n\rm PN}$ and 
${\cal \tilde L}^{\first}_{>n\rm PN}$ will 
be the same order as the ratio between 
${\cal \tilde L}^{\zeroth}_{n\rm PN}-{\cal L}^{\zeroth}_{0\rm PN}$ and 
${\cal \tilde L}^{\zeroth}_{>n\rm PN}$. 
Based on this assumption, we extrapolate the known results to estimate 
the unknown ${\cal \tilde L}^{\first}_{>n\rm PN}$. 

\section{The dephasing due to the averaged dissipative portion 
of the second-order self-forces}
\label{sec:N-PN}
The main focus of this section is to evaluate the dephasing 
from the averaged dissipative part of the second-order self-forces 
in E(I)MRIs, based on the idea 
of the extrapolation proposed in the preceding section. 
In this paper, we measure the magnitude of the dephasing 
due to higher-order corrections to the energy flux
$\delta{\cal L}$  by 
\begin{equation}
\label{delta-$n$PN}
\Delta \Phi[{\cal L},\delta{\cal L}] := 
-2 \int_{x_{\rm{ISCO}}}^{x_0} dx \frac{x^{3/2} {E}'{}^{\zeroth} (x) }{\mtot}
\left|  \frac{1}{ {\cal L}(x)+\delta {\cal L}(x)}-\frac{1}{{\cal L}(x)}
  \right|.
\end{equation}
Here the absolute value of the difference of fluxes is taken in 
the integrand. 
The factor 
$({\cal L}+\delta {\cal L})^{-1}-{\cal L}^{-1}$
can change its signature in the domain of the integral 
for some parameter region of E(I)MRIs. 
Therefore, if we do not take the absolute value, 
there might be an accidental cancellation 
between positive and negative contributions, 
and the order of magnitude of the evaluated $\Delta\Phi$ 
can largely deviate from what we really want to measure. 
Notice that, even if $\Delta\Phi$ defined without taking the absolute value 
in the integrand strictly vanishes,  the deviation in the GW waveform 
is still detectable. 
Therefore, to avoid this possible underestimate of 
the dephasing, we take the absolute value of the difference. 

As a general remark, we would like to mention the 
mass dependences of $\Phi[{\cal L}^{\zeroth}_{\rm ref}]$,
the phase for a certain reference flux ${\cal L}^{\zeroth}_{\rm ref}$ 
which includes the leading PN terms in the test particle limit, and 
$\Delta\Phi[{\cal L}^{\zeroth}_{\rm ref},\delta {\cal L}^{\zeroth}]$ 
and 
$\Delta\Phi[{\cal L}^{\zeroth}_{\rm ref},
\delta {\cal L}^{\first}]$, 
the dephasings due to $\delta {\cal L}^{\zeroth}$ and 
$\delta {\cal L}^{\first}$, 
which are portions of the energy flux at the leading order and 
the next leading order in the mass ratio, respectively.  
From the integral in Eq.~\eqref{test-year}, which defines $x_0$, 
one can factor out $\MBH/\nu$ since
\begin{equation} 
\label{scaling}
E'{}^{\zeroth}\propto \MBH \nu,\qquad  
{\cal L}^{\zeroth}_{\rm ref}  \propto \nu^2
\end{equation}
for a given $x$. 
Hence, we find that $x_0$ is a function of $\MBH/\nu$. 
Substituting $\delta{\cal L}^{\zeroth}\propto \nu^2$ and 
$\delta{\cal L}^{\first}\propto \nu^3$ 
together with the relations in Eq.~\eqref{scaling} into the 
definitions of 
$\nu\times \Phi[{\cal L}^{\zeroth}_{\rm ref}]$, 
$\nu\times \Delta\Phi[{\cal L}^{\zeroth}_{\rm ref},
\delta {\cal L}^{\zeroth}]$ 
and 
$\Delta\Phi[{\cal L}^{\zeroth}_{\rm ref} ,\delta{\cal L}^{\first}]$, 
we find that their mass dependences remain only through $x_0$. 
Therefore, one can conclude that 
$\nu\times \Phi[{\cal L}^{\zeroth}_{\rm ref}]$, 
$\nu\times \Delta\Phi[{\cal L}^{\zeroth}_{\rm ref},\delta {\cal L}^{\zeroth}]$  and
$\Delta\Phi[{\cal L}^{\zeroth}_{\rm ref},{\cal \tilde L}^{\first}]$ 
depend on $\MBH$ and $\nu$ only through the combination 
$\MBH / \nu$ for a large mass ratio $\nu\ll 1$ 
for a fixed observation period before the plunge.

\subsection{The error caused by the post-Newtonian truncation in the
  test particle limit}
\label{subsec:N-PN}
%
Before we evaluate the dephasing due to the 
yet-unknown part of the energy flux at the next leading order in 
the mass ratio, 
we will assess the magnitude of the dephasing due to the 
PN truncation in the test particle limit 
$\Delta\Phi[{\cal L}^{\zeroth}_{n\rm PN},{\cal L}^{\zeroth}_{>n\rm PN}]$
in this subsection and that coming from the known PN terms 
$\Delta \Phi[{\cal L}^{\zeroth}_{\num}
+{\cal \tilde L}^{\first}_{n\rm PN}, 
{\cal \tilde L}^{\first}_{>n\rm PN}]$ 
in the succeeding subsection.  

Here, we study the quantity 
$\Delta\Phi[{\cal L}^{\zeroth}_{n\rm PN},{\cal L}^{\zeroth}_{>n\rm PN}]$ 
to examine the PN convergence in terms of the phase error for various 
E(I)MRIs. 
We show the results only for the 
exponential resummed flux because the phase 
for the PN Taylor flux becomes ill defined for several 
PN orders. (The hybrid flux is identical to the exponential resummed one
by definition in the test particle limit.)



\begin{figure}[tbp]
\begin{center}  
\hspace{-3cm}
    \qquad \qquad 
    \includegraphics[width=6cm, clip, angle = -90]
    {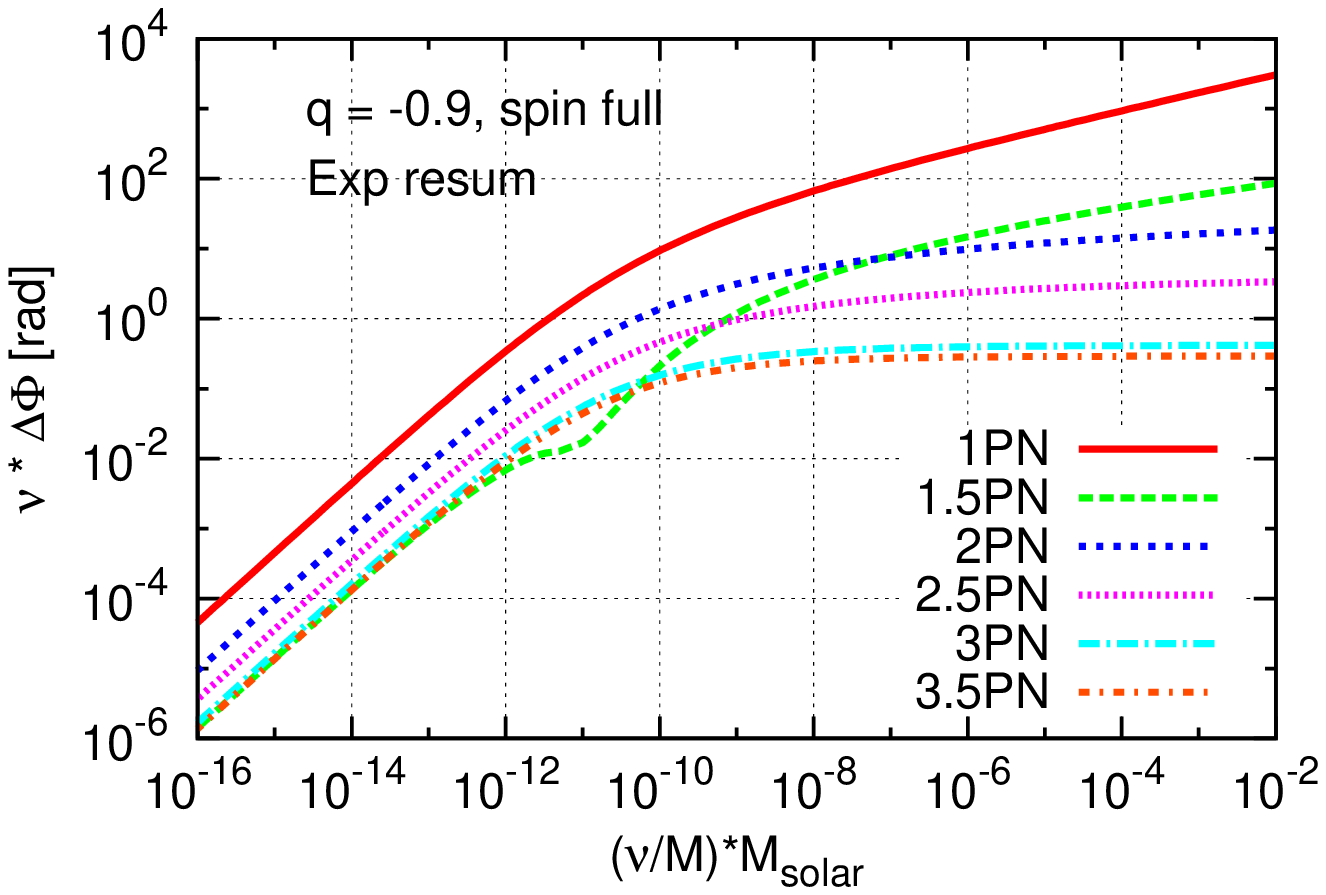}
    \includegraphics[width=6cm, clip, angle = -90]
    {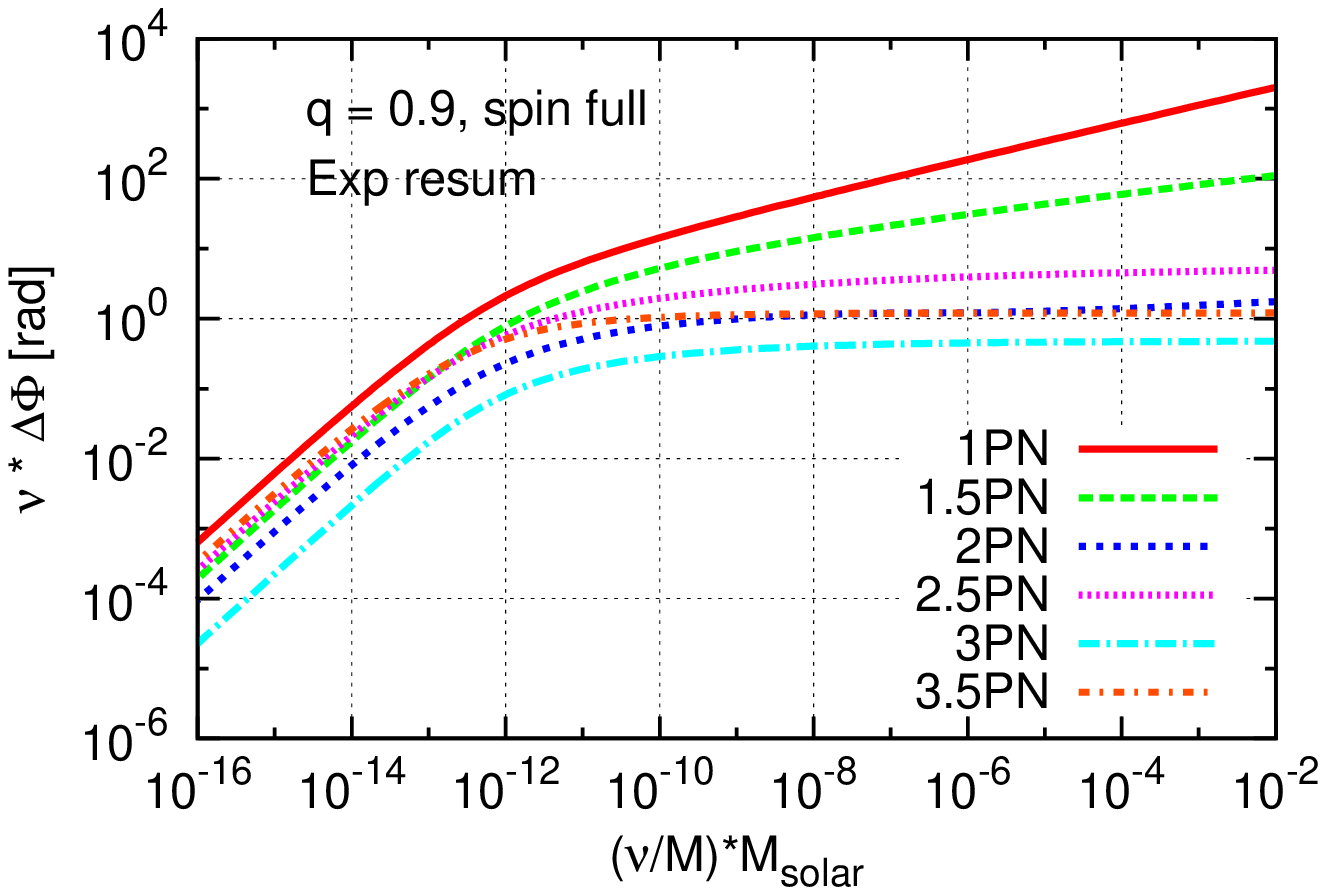}
 \end{center}
 \vskip -\lastskip \vskip -3pt
   \caption{
The dephasing due to the higher PN corrections in the test particle
 limit $\nu \Delta\Phi[{\cal L}^{\Exp \zeroth}_{n\rm PN},
{\cal L}^{\Exp \zeroth}_{>n\rm PN}]$ as a function of 
$\nu/\mtot$ for $q = \pm 0.9$. 
}
  \label{fig:PNconvergenceTP} 
\end{figure}%

In Fig.~\ref{fig:PNconvergenceTP} 
we plot 
$\nu \Delta\Phi[{\cal L}^{\Exp \zeroth}_{n\rm PN},
{\cal L}^{\Exp \zeroth}_{>n\rm PN}]$ as a function of 
$\nu/\mtot$ for $q = \pm 0.9$. 
The main message of these plots is that 
the PN convergence is almost uniform, except for the case of 
large black hole spin for large $\nu/\mtot$. 
When $q$ is large, ISCO becomes close to the light-ring radius 
where PN convergence becomes very slow. 
Even in that case, for lower PN orders 
the convergence is rather smooth for large $\nu/\mtot$. 
This is because up to the 2 PN order, the phase correction is dominated by 
the lower bound of the $x$-integral in Eq.~\eqref{delta-$n$PN}, 
if the initial separation of the binary 
at the time 1 yr before the plunge is sufficiently large. 
The reason we focus on large $\nu/M$ is that, 
as will be discussed in Append.~\ref{sec:post-PN}, 
the dephasing from the known PN terms is already well suppressed 
below 1 rad if $\nu/M$ is small enough.
Thus, the PN convergence for small $\nu/M$ does not 
affect the following discussion.

Notice also that $x_0$ becomes smaller and smaller for larger $\nu/\mtot$. 
The 2.5PN case is marginal, in the sense that 
the entire range of $x$ contributes almost equally. By contrast,  
the contribution near the ISCO dominates for the corrections at the 
3 PN order or higher. 
Although the PN convergence 
is not clearly seen for $q>0.9$, we should note that 
the phase error is never extremely enhanced for some particular 
post-Newtonian order, largely exceeding the values for the 3 PN or 
3.5PN order, upon which we mainly focus in the following discussion.

\subsection{The expected dephasing from the unkown higher-order 
post-Newtonian terms}
\label{sec:4-mass}
Now we move on to our main issue:
evaluating the dephasing due to the yet-unknown higher -order PN terms, 
$\Delta \Phi[{\cal L}^{\zeroth}_{\num}
+{\cal \tilde L}^{\first}_{3.5\rm PN}, 
{\cal \tilde L}^{\first}_{>3.5\rm PN}]$, 
based on the idea for the extrapolation 
proposed in Sec.~\ref{sec:8PN}. 
The scaling argument will tell us that 
the dephasings coming from the respective PN terms at the leading order 
in the mass ratio will be roughly proportional to the corresponding 
next-leading-order dephasings. Namely, 
\begin{equation}
\label{ratio}
\Delta \Phi[{\cal L}^{[0]}_{0 {\rm PN}}
,{\cal \tilde L}^{[0]}_{n\rm PN}-  {\cal L}^{[0]}_{0 {\rm PN}}]:
\Delta \Phi[{\cal \tilde L}^{[0]}_{n\rm PN},
{\cal \tilde L}^{[0]}_{>n\rm PN}] \approx
\Delta \Phi[{\cal \tilde L}^{[0]}_{\rm full},
{\cal \tilde L}^{[1]}_{n\rm PN} ]:
\Delta \Phi[{\cal L}^{\zeroth}_{\num}
+{\cal \tilde L}^{\first}_{n \rm PN}, {\cal \tilde L}^{[1]}_{>n\rm PN}]
\end{equation}
will hold. 
Recall that ${\cal \tilde L}^{[0]}_{n\rm PN}$ consists 
of the terms at the leading order in mass ratio 
up to the same $n$PN order that is available in 
${\cal \tilde L}^{[1]}_{n\rm PN}$.

Here, we have three basic options in our choice of flux: 
the Taylor flux, the exponential resummed flux and the hybrid flux. 
(Recall that the hybrid flux is defined as such that is 
identical to the exponential resummed flux at the 
leading order in the mass ratio.)
On one hand, the Taylor flux is problematic since it can be negative before 
reaching ISCO for some PN orders, say 2.5PN, 
with a modelate value of the spin parameter as we mentioned earlier.  
In such cases the phase before the plunge is not well defined.
On the other hand, 
the difference between the exponential resummed flux and 
the hybrid flux is negligible. 
Therefore, we here consider the hybrid flux only. 



\begin{figure}[tbp]
\begin{center}  
\hspace{-3cm}
    \qquad \qquad 
    \includegraphics[width=6cm, clip, angle = -90]
    {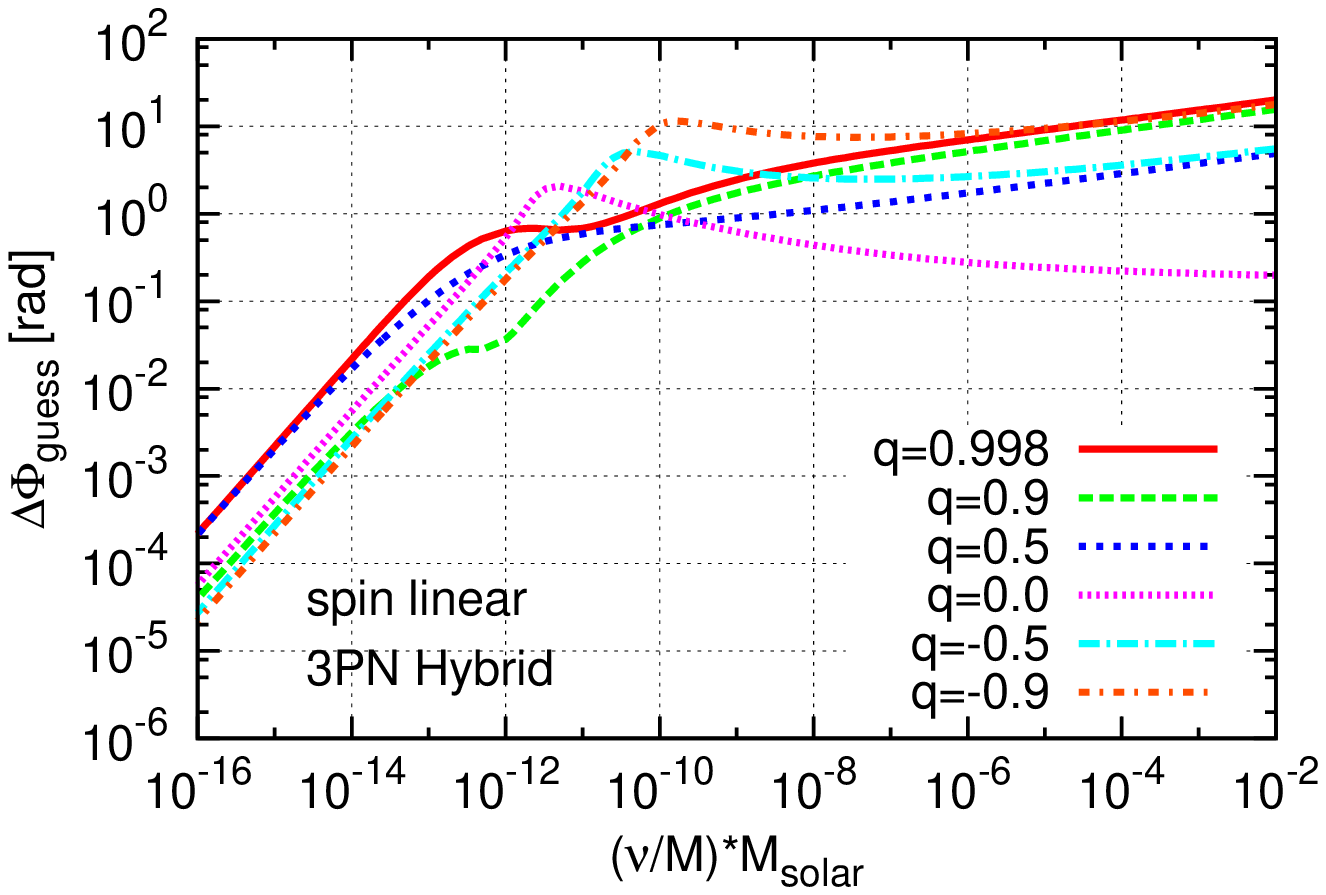}
    \includegraphics[width=6cm, clip, angle = -90]
    {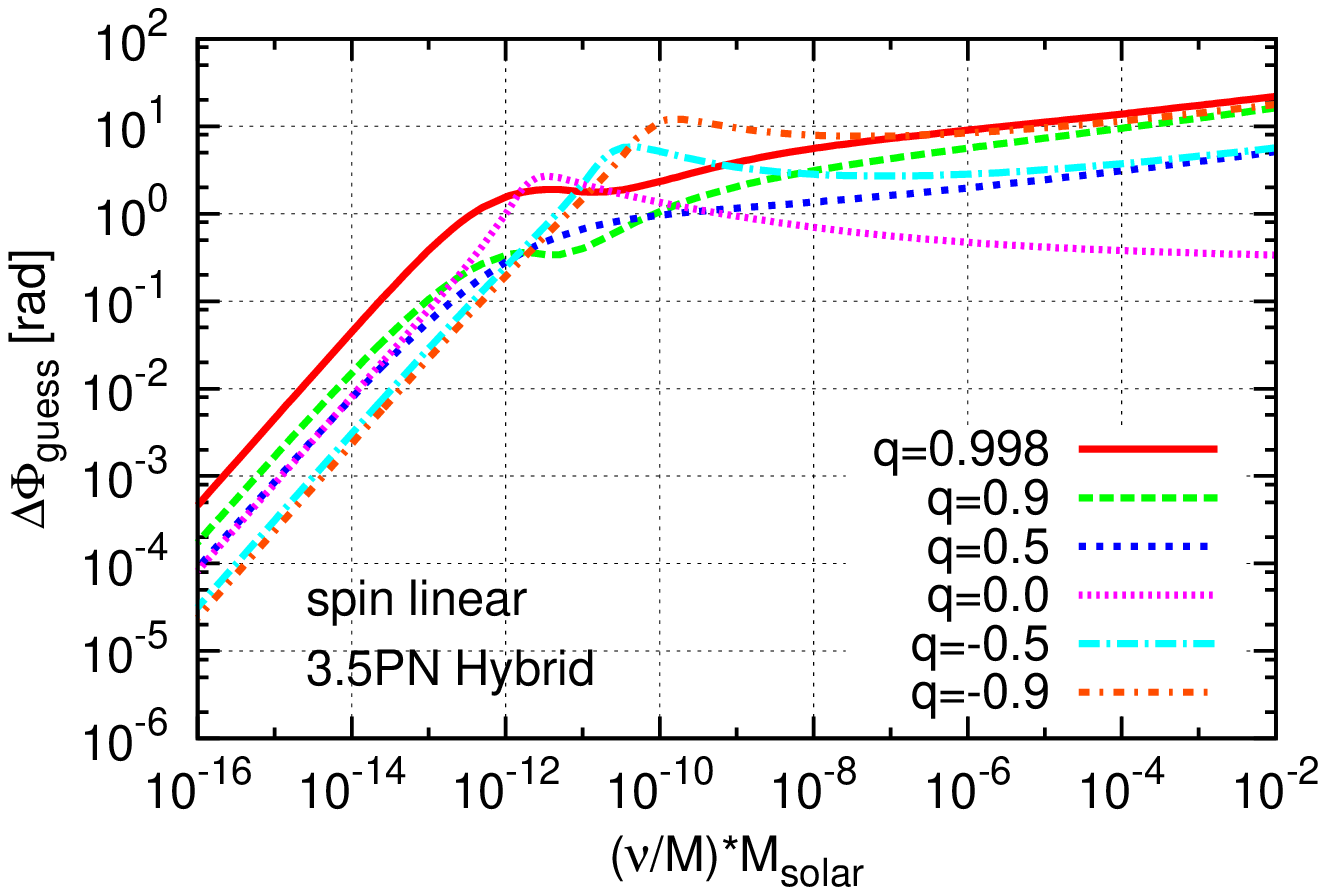}
 \end{center}
 \vskip -\lastskip \vskip -3pt
   \caption{
The expected residual dephasing 
$\Delta \Phi_{\estimate}[{\cal L}^{\zeroth}_{\num}
+{\cal \tilde L}^{\first}_{n\rm PN}, 
{\cal \tilde L}^{\first}_{>n\rm PN}]$ 
caused by the unknown part of the averaged dissipative second-order
 self-forces. 
For $n=3$, the 3.5PN known spin-independent 
part of the flux at the next leading order in the mass ratio 
is treated as unknown. 
   }
     \label{fig:pT-vary-q}
\end{figure}%
Using Eq.~\eqref{ratio}, the residual dephasing estimated by 
\begin{equation}
\Delta \Phi_{\estimate}[{\cal L}^{\zeroth}_{\num}
+{\cal \tilde L}^{\first}_{n\rm PN}, 
{\cal \tilde L}^{\first}_{>n\rm PN}]
:={
\Delta \Phi[{\cal \tilde L}^{[0]}_{n\rm PN},
{\cal \tilde L}^{[0]}_{>n\rm PN}]
\over 
\Delta \Phi[{\cal L}^{[0]}_{0\rm PN}
,{\cal \tilde L}^{[0]}_{n\rm PN}-{\cal L}^{[0]}_{0\rm PN}]
}
\Delta \Phi[{\cal \tilde L}^{[0]}_{ {\rm full} },
{\cal \tilde L}^{[1]}_{n\rm PN}], 
\label{residualdephasing}
\end{equation}
with $n=3$ and $n=3.5$, is depicted in Fig.~\ref{fig:pT-vary-q}. 
The trend of the estimated dephasing is independent of 
the truncated PN order. 
The point of Fig.~\ref{fig:pT-vary-q} is that the residual dephasing 
is rather suppressed over the whole range of the binary parameters.
Indeed, the residual dephasing is at most about 10 rad in 
the case of 3 PN truncation. 
In the case of 3.5PN truncation, the maximum value of the 
residual dephasing is bigger by a factor of two or so. 
Since the plot is given in radians, the value 
must be divided by 2$\pi$ to translate it into the number of cycles. 
From the above results, one may say that 
the residual dephasing due to the yet-unknown PN corrections 
at the next leading order in the mass ratio 
$\Delta \Phi_{\estimate}[{\cal L}^{\zeroth}_{\num}
+{\cal \tilde L}^{\first}_{n\rm PN}, 
{\cal \tilde L}^{\first}_{>n\rm PN}]$
is not negligible, adopting one cycle as the criterion 
for the significant dephasing.

The curves in Fig.4 are, roughly speaking, broken power-law curves.  
The shallower slope on the right corresponds to the range   
in which the initial separation of the binary is sufficiently 
large. In this case the suppression due to the 
mass ratio in the higher-order correction is compensated for by 
the longer duration spent by the orbit in a given frequency band. 
By contrast, the steeper slope on the left corresponds to the region  
in which the initial separation is small.
In this case, the total number of cycles is cut off by the 
observation period, and hence it does not increase for smaller $\nu/M$. 
As a result, the dephasing rapidly decreases for smaller $\nu/M$ 
below the critical value determined by the observation period.

In Fig.~\ref{fig:pT-vary-q}, we also see a tiny bump around 
$\nu M_{\odot}/\mtot \approx 10^{-10}-10^{-11}$ for some values of $q$. 
The reason why this bump appears at this position 
can be understood from the observation that 
${\cal \tilde L}^{\zeroth}_{n\rm PN}-{\cal L}^{\zeroth}_{0\rm PN}$ 
crosses zero within the domain of the integral, $x_0 < x < x_{\rm ISCO}$ 
when the bump appears.
When ${\cal \tilde L}^{\zeroth}_{n\rm PN}-{\cal L}^{\zeroth}_{0\rm PN}$ 
crosses zero, 
the factor in the denominator of Eq.~\eqref{residualdephasing}
$\Delta \Phi[{\cal L}^{[0]}_{0\rm PN}
,{\cal \tilde L}^{[0]}_{n\rm PN}-{\cal L}^{[0]}_{0\rm PN}]$
suffers from more or less accidental suppression 
even though we take the absolute value 
in the integrand of $\Delta\Phi$ in Eq.~\eqref{delta-$n$PN}.
We denote the value of $x$ at which the flux 
${\cal \tilde L}^{\zeroth}_{n\rm PN}-{\cal L}^{\zeroth}_{0\rm PN}$ 
crosses zero by $x_{\rm cross}$. 
Then, if $x_0\ll x_{\rm cross}$ or $x_0\gg x_{\rm cross}$,  
this suppression does not produce much effect on the estimate of 
$\Delta \Phi[{\cal L}^{[0]}_{0\rm PN},
{\cal \tilde L}^{[0]}_{n\rm PN}-{\cal L}^{[0]}_{0\rm PN}]$. 
Therefore, the suppression becomes significant only 
for $x_0\approx x_{\rm cross}$. 
The value of $x_{\rm cross}$ is rather close to the value at ISCO, 
$x_{\rm ISCO}$, but typically not extremely close to it. 
Hence, $x_0\approx x_{\rm cross}$ occurs when $x_0$ 
is neither 
extremely close to ISCO nor very small like $x_0\ll 1$, 
which corresponds to the break of the curves at 
$\nu M_{\odot}/\mtot\approx 10^{-10}-10^{-11}$. 
Since now we find that this tiny bump is to be attributed to 
an accidental zero in 
${\cal \tilde L}^{\zeroth}_{n\rm PN}-{\cal L}^{[0]}_{0\rm PN}$, 
this bump would be regarded as an artificial feature. 
If it is fair to remove the bumps from Fig.~\ref{fig:pT-vary-q}, 
we will find that the significantly large 
dephasing (in the sense of exceeding one cycle) 
will be expected only for $\nu M_{\odot}/\mtot\gtrsim 10^{-9}$.

\subsection{Unknown dephasing expected when we know 
the lower-PN-order nonlinear spin-dependent terms in the energy flux}
\label{subsec:N-spin}
In the preceding subsection, we find that the dephasing caused by 
the unknown PN higher-order terms at the next leading order in mass
ratio can be $O(10)$ rad or more. However, these unknown terms in 
the flux include nonlinear spin terms at the 3 PN or lower order 
and all spin-dependent terms at the 3.5PN order, 
and hence we have excluded the corresponding terms 
from ${\cal \tilde L}^{[0]}_{n\rm PN}$, which appears 
in the expression Eq.~\eqref{residualdephasing}. 
In Fig.~\ref{fig:pT-vary-q}, relatively large dephasing is 
observed for large $|q|$, for which the nonlinear spin terms 
will be important. 

Here we consider how the estimate presented in the preceding subsection 
is modified once we obtain all the spin-dependent terms up to 3 PN or 3.5PN 
order. 
What we need to evaluate is the expression 
obtained by removing $\tilde{}$ from \eqref{residualdephasing}, \ie,
\begin{equation}
\Delta \Phi_{\estimate}[{\cal L}^{\zeroth}_{\num}
+{\cal L}^{\first}_{n\rm PN}, 
{\cal L}^{\first}_{>n\rm PN}]
:={
\Delta \Phi[{\cal L}^{[0]}_{n\rm PN},
{\cal L}^{[0]}_{>n\rm PN}]
\over 
\Delta \Phi[{\cal L}^{[0]}_{0\rm PN}
,{\cal L}^{[0]}_{n\rm PN}-{\cal L}^{[0]}_{0\rm PN}]
}
\Delta \Phi[{\cal L}^{[0]}_{\rm full},
{\cal \tilde L}^{[1]}_{n\rm PN}]. 
\label{residualdephasing2}
\end{equation}
Here, one may think that 
${\cal \tilde L}^{[1]}_{n\rm PN}$ in the factor 
$\Delta \Phi[{\cal L}^{[0]}_{n\rm PN},
{\cal \tilde L}^{[1]}_{n\rm PN}]$ should have also been replaced with 
${\cal L}^{[1]}_{n\rm PN}$.  
Since we do not have the expression for 
${\cal L}^{[1]}_{n\rm PN}$ 
at hand, we cannot perform this replacement. 
However, the difference between ${\cal L}^{[1]}_{n\rm PN}$ 
and ${\cal \tilde L}^{[1]}_{n\rm PN}$ will not be significant, 
because their dominant PN terms are common.



\begin{figure}[tbp]
\begin{center}  
\hspace{-3cm}
    \qquad \qquad 
    \includegraphics[width=6cm, clip, angle = -90]
    {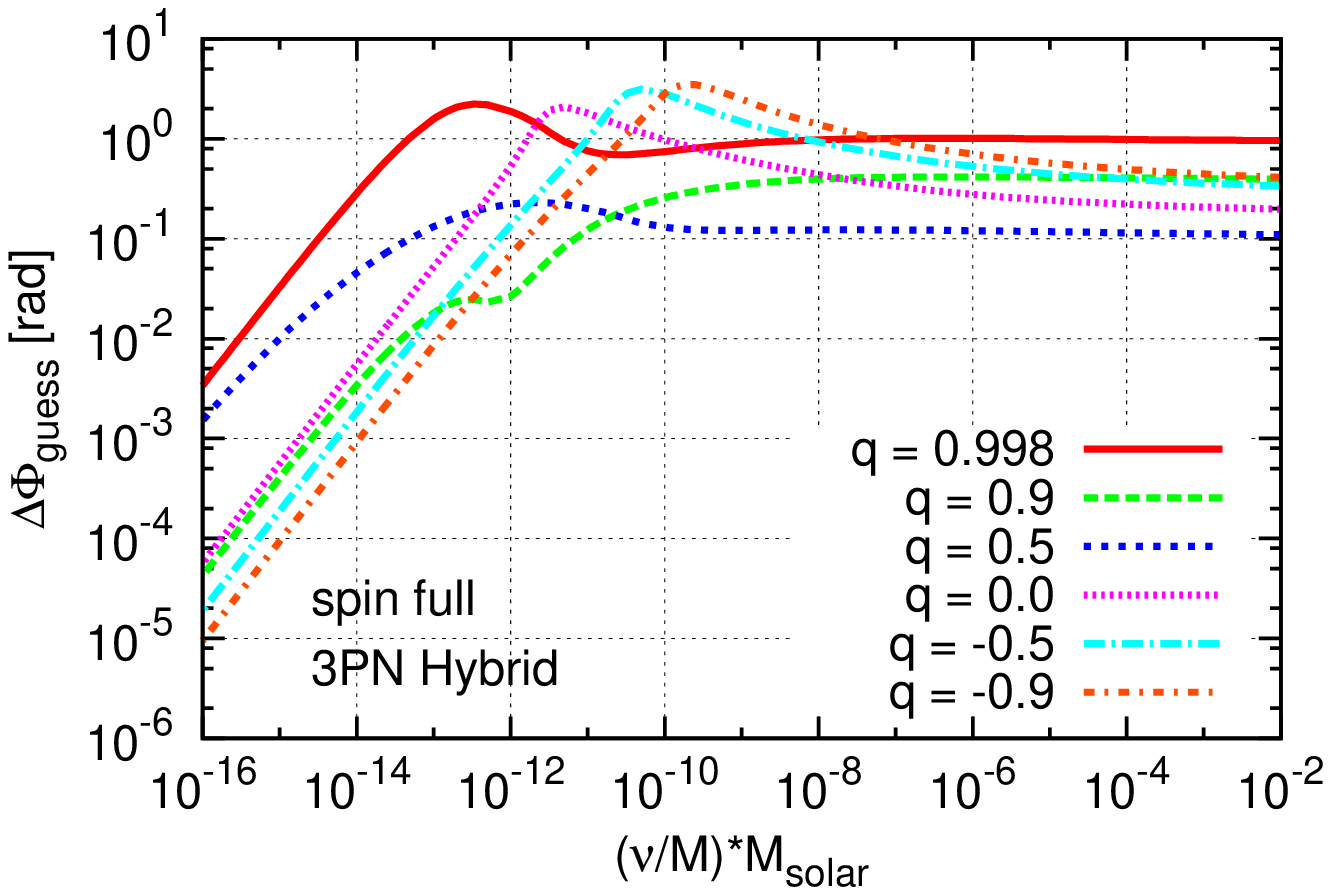}
    \includegraphics[width=6cm, clip, angle = -90]
    {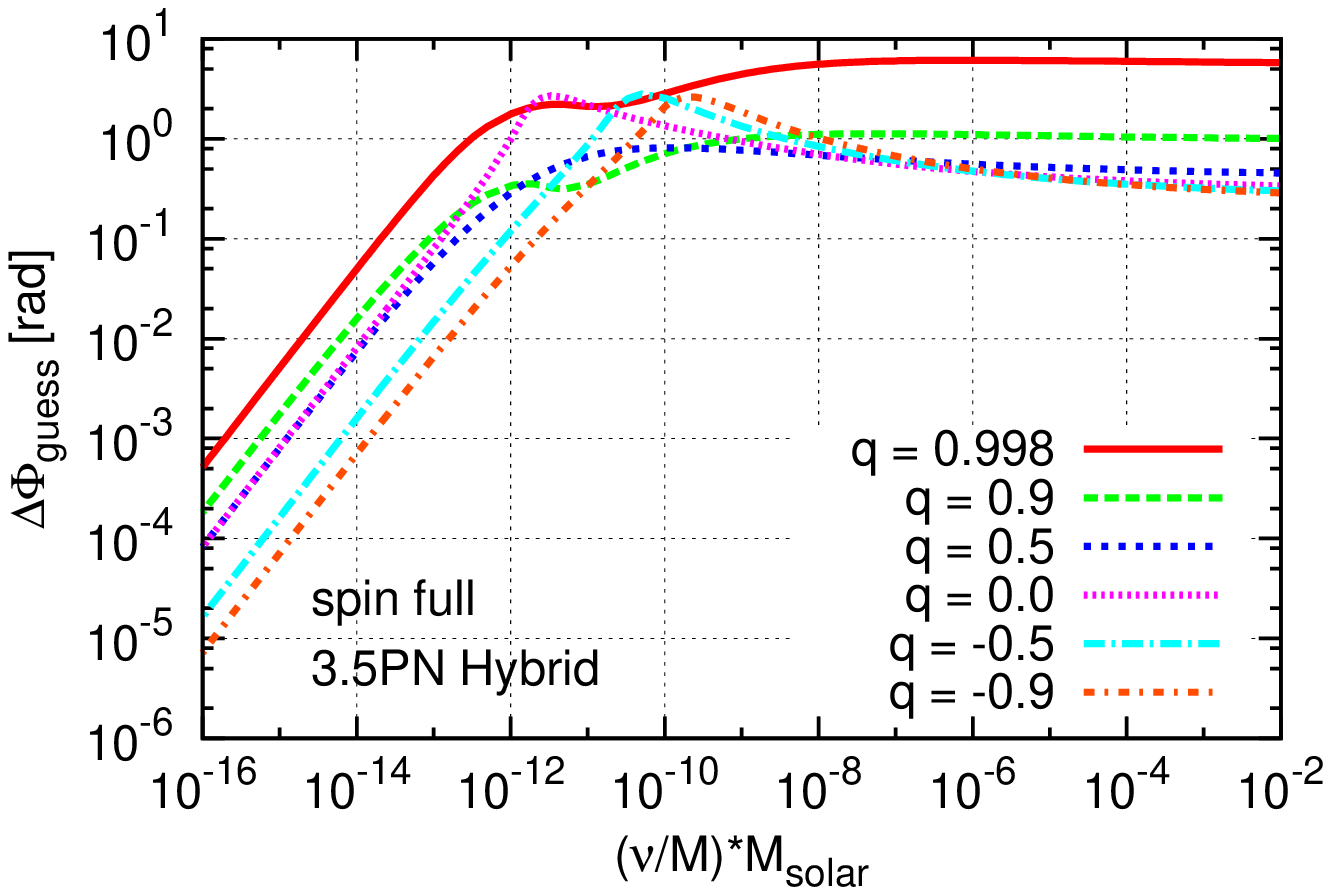}
 \end{center}
 \vskip -\lastskip \vskip -3pt
   \caption{The expected residual dephasing 
$\Delta \Phi_{\estimate}[{\cal L}^{\zeroth}_{\num}
+{\cal L}^{\first}_{n\rm PN}, 
{\cal L}^{\first}_{>n\rm PN}]$ when we assume that 
the spin-dependent terms at the lower PN orders are all known. 
}
     \label{fig:spin}
\end{figure}%

The resultant dephasing 
$\Delta \Phi_{\estimate}[{\cal L}^{\zeroth}_{\num}
+{\cal L}^{\first}_{n\rm PN}, 
{\cal L}^{\first}_{>n\rm PN}]
$
is depicted in Fig.~\ref{fig:spin}.
In contrast to the Fig.~\ref{fig:pT-vary-q},
except for the near-extremal-spin case ($q = 0.998$) with 
3.5PN truncation ($n=3.5$), 
the residual dephasing for large $\nu/\mtot$ evaluated by using the hybrid
flux does not exceed 1 rad. 
The residual dephasing exceeds 1 rad only around the peak, 
which is likely to be an artifact due to the suppression 
in $\Delta \Phi[{\cal L}^{[0]}_{0\rm PN},
{\cal L}^{[0]}_{n\rm PN}-{\cal L}^{[0]}_{0\rm PN}]$ 
similarly to the previous case.  
The above result indicates that the dominant source of the 
error caused by the unknown part of the flux is in the 
spin-dependent terms at the lower PN orders up to the 3 PN or 3.5PN order. 
Namely, the knowledge of nonlinear spin-dependent terms in the energy flux is 
crucial for reducing the uncontrolled dephasing caused by the averaged 
dissipative second-order self-forces.
In addition, once we obtain these spin dependent terms 
of the energy flux, the exponential resummation will improve 
its accuracy to a level almost sufficient for the detection 
of E(I)MRIs in the whole interesting parameter region.

\section{Summary and conclusion}
\label{sec:summary}

To evaluate the second-order self-forces in the context of the black
hole perturbation is a challenging issue motivated by the goal to extract 
information from extreme (intermediate) mass-ratio inspirals [E(I)MRIs]. 
In this paper we have assessed the dephasing 
of the GW waveform from quasicircular E(I)MRIs, 
caused by the averaged dissipative part of the second-order self-forces, 
focusing on giving an order-of-magnitude estimate of the influence of 
the yet-unknown higher-order post-Newtonian (PN) corrections. 
Using the balance argument, the dephasing is related to the correction to the 
the emitted energy flux. Although it will still require further 
efforts to calculate the second-order self-forces 
using the black hole perturbation,  
they are partially already known from the standard PN calculation. 
We gave an estimate of how much dephasing will be 
caused by the yet-unknown higher-order PN terms 
for the last 1 yr of inspiral before the plunge, 
exhaustively exploring the whole possible parameter 
region of E(I)MRIs.

To give a guess for the yet-unknown higher-order PN terms, 
we first introduced a simple resummation method 
for the energy flux, which we call the exponential resummation. 
This is simply obtained by exponentiating the energy flux and truncating 
it at the known PN order in the exponent.  
This resummation has three merits: 
it ensures the positivity of the energy flux, 
it accelerates the PN convergence, 
and it can be applicable if we just know the PN Taylor flux.  
Since the PN Taylor flux in the test particle limit with large $q$ 
is known to be negative outside the innermost stable circular orbit
(ISCO) at several PN orders, 
the total phase before the plunge becomes ill defined. 
Hence, we cannot discuss the amplitude of the higher PN corrections 
by using the PN Taylor flux. The exponential resummation solves this 
problem and improves the PN convergence in terms of the total phase before 
the plunge. When we discuss the finite mass corrections to the energy flux, 
we can combine the idea of the exponential resummed flux 
with the known exact energy flux in the test particle limit.  
We also proposed such a phenomenological energy flux, which we 
call the hybrid flux. 

To examine the dephasing from the unknown part of the averaged dissipative 
second-order self-forces, we need some extrapolation. 
Using the brand-new 8 PN energy flux in 
the test particle limit\cite{Fujita2012:un} 
that includes all spin-dependent terms, 
we discovered that the order of the magnitude of the 
absolute values of the coefficients up to the 8 PN order 
approximately follows a simple scaling law, which is  
what we expect when the actual convergence radius of the PN expansion
is at the light ring radius, 
irrespective of the value of dimensionless spin parameter $q$.
Since there is no reason to expect that the energy flux diverges outside 
the light-ring radius even if we take into account the finite mass corrections 
for quasicircular E(I)MRIs, we are motivated to assume that the ratio between 
the magnitude of the terms at $O(\nu^0)$ and $O(\nu)$ 
at the same PN order will roughly stay independent of the PN order. 

Based on this assumption, we estimated 
the unknown portion of the energy flux  
that comes from the higher PN terms at the next leading order in 
the mass ratio and evaluated the residual dephasing due to them. 
We find that the residual dephasing 
may exceed one cycle for 
$\nu M_{\odot} / \mtot \gtrsim 10^{-10}$ 
and spin of the Kerr blackhole $|q| > 0.5$, 
assuming a 1 yr observation period. 
For some parameters, we found a little enhancement of 
the estimated dephasing at 
$\nu M_{\odot}/\mtot \approx 10^{-10}-10^{-11}$. 
Since this enhancement is likely to be attributed to a mild accidental 
cancellation in the factor in the denominator of the estimator of 
the residual dephasing that we adopted, it might be fair to remove 
the bump that arises for this reason.   
Then, the residual dephasing exceeds one cycle 
only for relatively large $O(\nu/\mtot)$ with $|q| > 0.5$. 
Even for rather extreme case like $q=0.998$ (the Thorne limit), 
the expected residual dephasing is at most a few cycles or so. 

In the unknown flux, nonlinear spin terms at lower PN orders are 
also included since they are not yet calculated in the context of 
PN approximation even at the leading order in the PN expansion. 
As large residual dephasing is expected only for large $|q|$, 
one may suspect that the residual dephasing might 
be dramatically reduced once we know all the nonlinear spin-dependent 
terms up to the 3 PN or 3.5PN order. 
Therefore we made analogous plots for the expected residual dephasing 
assuming that the nonlinear spin-dependent terms are also available.  
We found that the expected residual dephasing is further suppressed 
to be less than 1 rad for $|q|<0.9$, as we expected.  
Even in the extreme case with $q=0.998$, 
the expected dephasing becomes less than
1 rad if we use 3 PN truncation to evaluate the residual. 
We think the reason why 3.5PN truncation is worse is simply 
because, for large $q$ close to unity, ISCO is very close 
to the light-ring radius, and hence the PN convergence is very poor. 
Thus, the resulting estimate of the residual also fluctuates 
by a large order of magnitude in this limit. 
However, even if we rely on 3.5PN truncation, the expected 
dephasing is at most about one cycle for $q=0.998$.

Although a 1 yr observation period is assumed  
in all the above estimates, the results do not depend much on it. 
A change in the observation time varies 
the relation between $x_0$ and $\nu/\mtot$, and hence 
the curves in Figs.~\ref{fig:pT-vary-q} and 
\ref{fig:spin} 
are just shifted horizontally leftward. 
The expected dephasing becomes large only 
for large $\nu/\mtot$, but $x_0$ 
is already sufficiently small in this case. 
Therefore the effect of changing the observation period 
for large $\nu/\mtot$ appears only in Fig.~\ref{fig:pT-vary-q}, 
in which lower-order PN terms are contributing. 
In fact, one can see that the curves in Figs.~\ref{fig:spin} 
are already saturated for large $\nu/\mtot$.

To conclude, the residual dephasing 
caused by the unknown averaged dissipative part of the 
second-order self-forces is estimated to be at most a few cycles. 
What is more, this dephasing is mostly to be attributed to the lower-PN-order 
nonlinear spin-dependent terms. If the PN expansion to the 3 PN or 3.5PN 
order is completed including the dependence on the black hole spin, 
the residual is expected to be further reduced. 
We think that this conclusion will not change 
even if we are slightly underestimating the dephasing.

It should be stressed that we are not trying to claim that all the 
second-order terms can be neglected for detection.  
The known part of the averaged dissipative second-order self-forces 
might be necessary even just for detection. 
Moreover, when the parameter extraction from EMRIs is concerned, even a small 
dephasing in principle gives a bias. 
Hence, to what extent the higher-order corrections are needed 
crucially depends on the accuracies required by the physics 
that we wish to extract and also on the signal-to-noise ratio.

As a final remark, we should note 
that there is a possibility that we might be underestimating the residual 
dephasing by an order of magnitude, since the PN 
convergence is not very smooth, especially for positive large $q$.  
Even if our estimate turns out to be a good approximation 
of the real magnitude, 
we still have a chance to have a golden event with a large 
signal-to-noise ratio. 
In that case, one cycle or even one radian 
may not be a sufficient accuracy for the template waveforms. 
Then, the dissipative part of the second-order self-forces that cannot be 
captured by the PN expansion becomes necessary to extract the best 
physics from the observational data. 
Also we should mention that we focused on quasicircular E(I)MRIs 
in this paper, but our scaling argument will not apply 
anymore for significantly eccentric orbits, since 
there are many modes whose frequency exceeds the naive PN convergence 
radius that will be given by the absolute value 
of the first complex quasi-normal mode frequency. 
Therefore, the hybrid energy flux cannot be expected to remain a good 
approximation near the plunge for such systems. 
Hence, our analysis does not at all discourage 
the study on the averaged dissipative part of the second-order self-forces. 
Our claim is that it will be possible to perform nearly the best 
analyses of most (nearly) quasicircular E(I)MRIs 
without waiting for the full development of 
our knowledge about the averaged dissipative part of 
the second-order self-forces.


\acknowledgments
S.I. acknowledges the support of the Grant-in-Aid for JSPS Fellows, 
No. 24-4281.
R.F. is grateful for the support of the European Union FEDER funds, 
the Spanish Ministry of Economy and Competitiveness (Project No. 
FPA2010-16495 and No. CSD2007-00042) and the Conselleria 
d'Economia Hisenda i Innovacio of the Govern de les Illes Balears. 
R.F. also appreciates the warm hospitality at the Yukawa Institute 
for Theoretical Physics where part of this work was completed.
H.T. is supported by the Grand-in-Aid for Scientific Research 
(No 23540309), and T.T. is supported by the Grand-in-Aid for 
Scientific Research 
(No. 21111006, No. 21244033, No.24103001, and No. 24103006).
Finally, this work was supported by the Grant-in-Aid for the Global COE Program
``The Next Generation of Physics, Spun from Universality and Emergenc''
from the Ministry of Education, Culture, Sports, Science and Technology
of Japan. 

%

\appendix


\section{Dephasing from the known post-Newtonian terms}
\label{sec:post-PN}
\begin{figure}[tbp]
\hspace{-3cm}
   \begin{center}
    \includegraphics[width=0.36\hsize, clip, angle = -90]
    {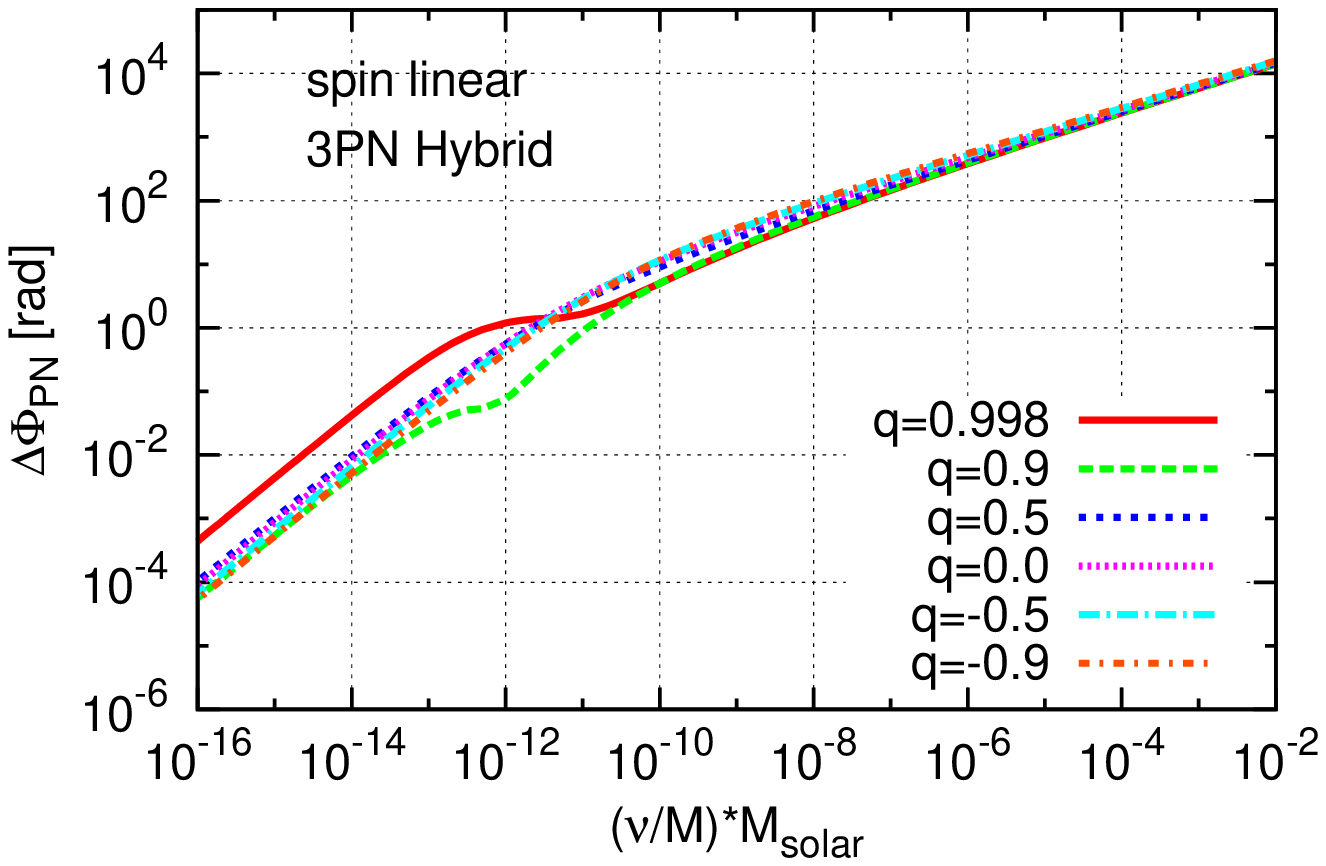}
  \end{center}
 \vskip -\lastskip \vskip -3pt
   \caption{The dephasing caused by the known terms up to the 
 3 PN order at the next leading order in the mass ratio, 
$\Delta\Phi_{\rm PN} [{\cal L}^{\zeroth}_{\num},{\cal \tilde 
 L}^{\hybrid\first}_{3\rm PN}]$, 
   for various dimensionless spin parameters $q$ of a Kerr black hole, 
   based on the 3 PN hybrid flux
   ~\eqref{hybrid}. 
   }
     \label{fig:vary-q}
\end{figure}

%
This appendix is dedicated to finding the dephasing 
from the available PN terms in the next leading order in the 
mass ratio.
In the bulk of our paper, 
we introduced three different types of PN flux formulas:  
Taylor, exponential resummed and hybrid, but 
there is no qualitative difference among 
$\Delta\Phi_{\rm PN} [{\cal L}^{\zeroth}_{\num}
,{\cal \tilde L}^{\T\first}_{n\rm PN}]$,
$\Delta\Phi_{\rm PN}
[{\cal L}^{\zeroth}_{\num},
{\cal \tilde L}^{\Exp\first}_{n\rm PN}]$ and
$\Delta\Phi_{\rm PN}
[{\cal L}^{\zeroth}_{\num},{\cal \tilde L}^{\hybrid\first}_{n\rm PN}]$, 
as long as we choose the exact numerical flux ${\cal L}^{\zeroth}_{\num}$
as the reference flux. 
This is because the significant difference among various fluxes 
appears only in the higher PN residual parts, such as those discussed in 
Sec.~\ref{sec:N-PN}.
Therefore, we show only $\Delta\Phi_{\rm PN}
[{\cal L}^{\zeroth}_{\num},{\cal \tilde L}^{\hybrid\first}_{3\rm PN}]$ 
in Fig.~\ref{fig:vary-q}.
We stress that what we show is the dephasing caused by
the terms not at the 3 PN order but up to the 3PN order. 
The main message of Fig.~\ref{fig:vary-q} 
is that the corrections to the phase of GWs 
due to the higher-order terms in the mass ratio 
are suppressed below 1 rad,  
even if we include the leading order of the PN expansion-
for a binary with $\nu M_\odot/\mtot<10^{-12}$, say, 
when the mass of the central Kerr black hole is greater than 
$3\times 10^6 M_\odot$ with the satellite mass fixed to 10$M_\odot$.
This holds irrespective of the value of the dimensionless spin 
parameter $q$. 
This plot clearly denies the naive statement that the effects of 
higher PN corrections should be important for a more massive 
central black hole because the satellite stays near the ISCO 
for a longer period. 



As for results in the literature for comparison, 
for a binary with $\MBH := 10^6 M_{\odot},~q = 0$ and 
$\nu = 10^{-5}$, Heurte and Gair\cite{Huerta:2008gb} 
reported 1.5 rad dephasing after 
the last 1 yr of inspiral based on the 2 PN Taylor flux,  
while Yunes~{\etal}\cite{Yunes:2010zj} 
reported 0.71~rad dephasing after the last 11.5 months of inspiral 
for the same binary parameters. 
Following our definition of the dephasing, 
we calculate $\Delta\Phi_{\rm PN}
[{\cal L}^{\zeroth}_{\num},{\cal \tilde  L}^{\hybrid\first}_{2\rm PN}]$ 
to find 1.46 rad, which we think is in good agreement with the previous 
results, within the variance due to the different definition of dephasing. 

\section{Dephasing due to the horizon absorption flux}
\label{sec:N-horizon}
In the bulk of our manuscript, 
we neglected the flux absorbed by the horizon. 
In the test particle limit, 
Tagoshi {\etal}\cite{Tagoshi:1997jy} and 
Yunes {\etal}\cite{Yunes:2010zj} have already shown 
that the energy flux absorbed through the horizon cannot be neglected, 
leading to a large dephasing, especially when the Kerr black hole 
has large spin. 
However, this does not mean that the corrections due 
to the horizon absorption flux at the 
next leading order in the mass ratio are also important. 
Naively, the standard PN formalism is not suitable for 
calculating the horizon absorption flux since 
the black hole horizon is beyond the reach of the 
standard PN expansion, 
although there is a direction to evaluate the horizon absorption flux 
by relating it with the tidal field around each 
black hole\cite{Alvi:2001mx,Poisson:2004cw}.  

\begin{figure}[tbp]
\begin{center}  
\hspace{-3cm}
\begin{minipage}[b]{.36\hsize}
   \begin{center}
    \includegraphics[width=\hsize,clip, angle = -90]
    {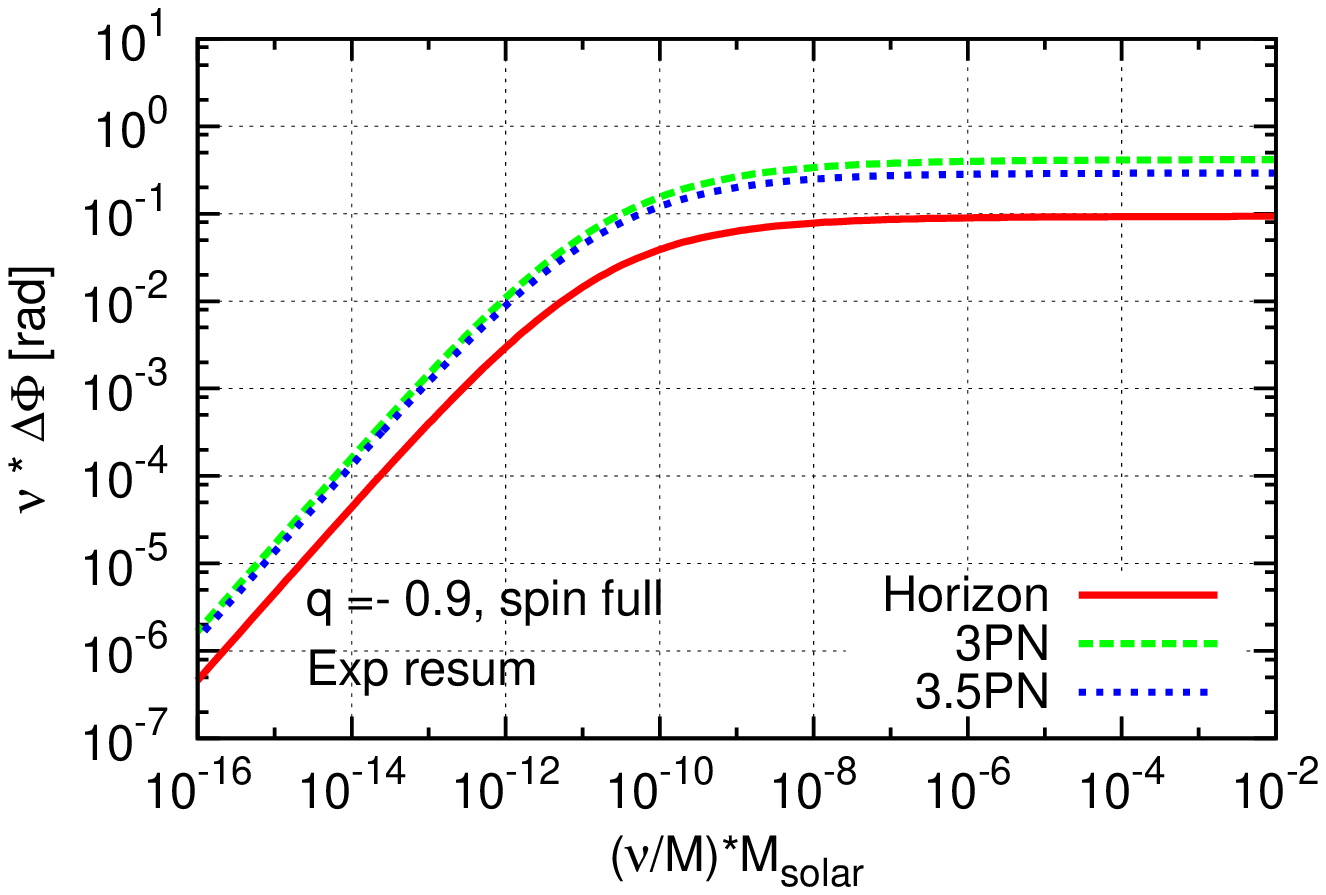}
  \end{center}
  \end{minipage}
  \qquad \qquad \qquad \qquad
 \begin{minipage}[b]{.36\hsize}
   \begin{center}
    \includegraphics[width=\hsize,clip, angle = -90]
    {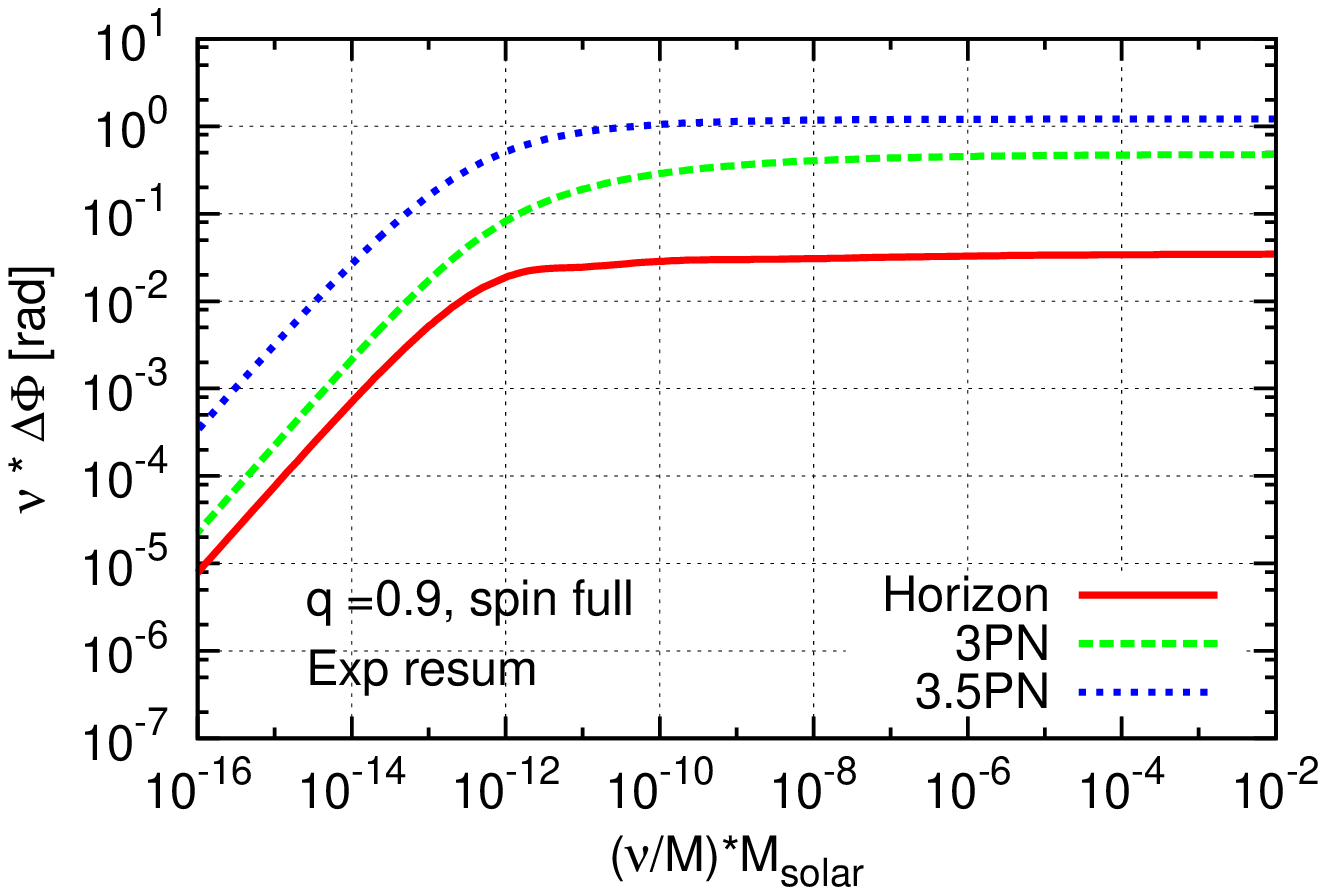}
  \end{center}
  \end{minipage}
 \end{center}
 \vskip -\lastskip \vskip -3pt
   \caption{
   The  dephasing 
   $\nu\Delta \Phi[{\cal L}^{[0]}_{\num}+{\cal L}^{[0,\rm H]}_{2.5\rm PN},
{\cal L}^{[0,\rm H]}_{>2.5\rm PN}]$
due to the horizon absorption flux 
and the dephasing $\nu \Delta\Phi[{\cal L}^{\Exp \zeroth}_{n\rm PN},
{\cal L}^{\Exp \zeroth}_{>n\rm PN}]$ due to the higher PN corrections,
for $n = 3$ and $n = 3.5$, and $q = \pm 0.9$.
	}
     \label{fig:horizon}
\end{figure}

In this appendix, we briefly 
address the dephasing caused by the flux absorbed through the horizon 
in the test particle limit to see what can be said about the 
dephasing due to the horizon absorption flux 
at the next leading order in the mass ratio. 
The flux absorbed through the horizon in the test particle limit
is expanded as\cite{Tagoshi:1997jy} 
\begin{eqnarray}\label{4-mass3}
{\cal L}^{[0,\rm H]}_{{n\rm PN}}(x)
&:=& \frac{32}{5}\nu^2 x^5 
\left\{ \left(- \frac{1}{4}q - \frac{3}{4}q^3   \right)x^{5/2} 
+ \left(- q - \frac{33}{16} q^3   \right)x^{7/2} 
+ O(x^{4}) 
\right\}~,  
\end{eqnarray}
and the exact numerical flux is also calculable\cite{Fujita:2004rb}.
We denote the contribution from the first term in the curly brackets 
in the above expression by 
${\cal L}^{[0,\rm H]}_{2.5\rm PN}$ and 
the remaining part of the horizon flux by 
${\cal L}^{[0,\rm H]}_{>2.5\rm PN}$, in a similar manner as before.
As we mentioned above, 
the horizon absorption flux at the next leading order in the mass ratio 
is still under development, but the part that requires the second-order 
metric perturbation should start with 1 PN order higher than the 
leading terms in the test particle limit, starting at the 3.5PN order. 
Hence, we adopt ${\cal L}^{[0,\rm H]}_{>2.5\rm PN}$, excluding 
${\cal L}^{[0,\rm H]}_{2.5\rm PN}$, as the counterpart of the unknown 
horizon absorption flux at the next leading order in the mass ratio. 
Under this consideration, 
we compare the dephasing due to the horizon absorption flux 
\begin{equation}
\Delta \Phi[{\cal L}^{[0]}_{\num}+{\cal L}^{[0,\rm H]}_{2.5\rm PN},
{\cal L}^{[0,\rm H]}_{>2.5\rm PN}]
\end{equation}
to that coming from 
the PN truncation $\Delta\Phi[{\cal L}^{\Exp \zeroth}_{n\rm PN},
{\cal L}^{\Exp \zeroth}_{>n\rm PN}]$ 
for $n=3$ or $n=3.5$ in the test particle limit.
The results are plotted in FIG.~\ref{fig:horizon} for $q = \pm 0.9$.


Fig.~\ref{fig:horizon} indicates that 
$\nu \Delta \Phi[{\cal L}^{[0]}_{\num}+{\cal L}^{[0,\rm H]}_{2.5\rm PN},
{\cal L}^{[0,\rm H]}_{>2.5\rm PN}]$
stays, at most, $O(10^{-1})$ rad for the entire E(I)MRI parameter region.
Indeed, it is, at most, about $1/10$ of 
$\nu \Delta\Phi[{\cal L}^{\Exp \zeroth}_{n\rm PN},
{\cal L}^{\Exp \zeroth}_{>n\rm PN}]$. 
For large $q$, the smallness of the effect of the horizon flux 
relative to the higher-order PN corrections 
can be understood from the work by Hughes\cite{Hughes:1999bq}, 
who numerically found that the energy flux absorbed by a Kerr black hole is 
,at best, ten times smaller than that emitted to infinity, with 
the aid of the fact that the PN expansion shows a poor convergence near 
the ISCO for large $q$. 
Thus, we expect that the residual dephasing due to the horizon absorption flux 
will be minor compared with the higher PN flux to infinity, and also 
in the next leading order in the mass ratio.


\end{document}